\documentclass[a4paper,12pt]{article}
\pdfoutput=1
\usepackage{jheppub,fancyhdr,graphicx,epstopdf,color,amsmath,cases,slashed,hyperref,multirow}
\usepackage{cancel}
\usepackage{float}
\usepackage{enumerate}
\usepackage{epsfig}
\usepackage{amsmath} 
\usepackage{amssymb} 
\usepackage{wasysym} 
\usepackage{mathrsfs} 
\usepackage{graphicx} 
\usepackage{amsfonts} 
\usepackage{amsbsy} 
\usepackage{amscd}
\usepackage{slashed} 
\usepackage{multirow} 
\usepackage{color}
\definecolor{myred}{rgb}{0.6,0,0} 
\definecolor{myblue}{rgb}{0,0.2,0.4}
\definecolor{mygreen}{rgb}{0,0.9,0.1}
\definecolor{hc}{rgb}{.9,0.1,0.7}
\definecolor{hcout}{rgb}{.9,0.7,0.9}
\definecolor{Orange}{rgb}{1.,0.65,0.}

\makeatletter

\newcommand{\fmslash}[2][0mu]{%
  \mathchoice
    {\fmsl@sh\displaystyle{#1}{#2}}%
    {\fmsl@sh\textstyle{#1}{#2}}%
    {\fmsl@sh\scriptstyle{#1}{#2}}%
    {\fmsl@sh\scriptscriptstyle{#1}{#2}}}
\newcommand{\fmsl@sh}[3]{%
  \m@th\ooalign{$\hfil#1\mkern#2/\hfil$\crcr$#1#3$}}
\makeatother

\newcommand{\lsim}{{\;\raise0.3ex\hbox{$<$\kern-0.75em\raise-1.1ex\hbox{$\sim$}}\;}}
\newcommand{\gsim}{{\;\raise0.3ex\hbox{$>$\kern-0.75em\raise-1.1ex\hbox{$\sim$}}\;}}

\newcommand{\met}{{\slashed{E}_T}}

\usepackage{array}
\newcolumntype{C}[1]{>{\centering\arraybackslash$}p{#1}<{$}}

\usepackage{bm} 
\newcommand{\be}{\begin{equation}}
\newcommand{\ee}{\end{equation}}
\newcommand{\bes}{\begin{equation*}}
\newcommand{\ees}{\end{equation*}}
\newcommand{\bea}{\begin{eqnarray}}
\newcommand{\eea}{\end{eqnarray}}
\newcommand{\beas}{\begin{eqnarray*}}
\newcommand{\eeas}{\end{eqnarray*}}
\newcommand{\snu}{\tilde{\nu}_{R}}
\newcommand{\stau}{\tilde{\tau}_{R}}
\newcommand{\smu}{\tilde{\mu}_{R}}
\newcommand{\sel}{\tilde{e}_{R}}
\newcommand{\neu}{\tilde{\chi}_{1}^{0}}
\newcommand{\staul}{\tilde{\tau}_{1}}
\title{Long-lived stau, sneutrino dark matter and right-slepton spectrum} 

\preprint{IPPP/18/44, LAPTH-021/18, HRI-RECAPP-2018-004}

\author[a]{Shankha Banerjee}
\author[b]{\!\!, Genevi\`{e}ve B\'{e}langer}
\author[c]{\!\!, Avirup Ghosh}
\author[c]{\!\!, Biswarup Mukhopadhyaya}

  \affiliation[a]{Institute for Particle Physics Phenomenology, Department of Physics, Durham University, Durham DH1 3LE, United Kingdom}
  \affiliation[b]{Universit\'{e} Grenoble Alpes, USMB, CNRS, LAPTh, F-74000 Annecy, France}
  \affiliation[c]{Regional Centre for Accelerator-based Particle Physics,
Harish-Chandra Research Institute, HBNI,
Chhatnag Road, Jhunsi, Allahabad - 211 019, India} 
\emailAdd{shankha.banerjee@durham.ac.uk}
\emailAdd{belanger@lapth.cnrs.fr}
\emailAdd{avirupghosh@hri.res.in}
\emailAdd{biswarup@hri.res.in}

\abstract{The minimal supersymmetric (SUSY) standard model (MSSM) augmented by right chiral sneutrinos
may lead to one such sneutrino serving as the lightest supersymmetric particle and a
non-thermal dark matter candidate, especially if neutrinos have Dirac masses only. In such cases,
if the lightest MSSM particle is a stau, the signal of SUSY at the LHC consists
in stable  charged tracks which are distinguishable from backgrounds through their
time delay between the inner tracker and the muon chamber. We show how to determine
in such scenarios the mass hierarchy between the lightest neutralino and right sleptons of the first
two families. The techniques of neutralino reconstruction, developed in earlier works, are
combined with the endpoint of the variable $M_{T2}$ in smuon (selectron) decays for this purpose.
We show that one can thus determine the  mass hierarchy for smuons (selectrons) and neutralinos
up to 1 TeV, to the level of 5-10\%.}


\begin{document}
\maketitle

\newpage

\section{Introduction} \label{sec:intro}

A persistent set of observations, encompassing things as diverse as galactic rotation~\cite{Sofue:2000jx}, 
anisotropy of the cosmic microwave background radiation (CMBR)~\cite{Ade:2015xua} and gravitational lensing 
effect around the tail of bullet clusters~\cite{Markevitch:2001ri}, have established that 23.8$\%$ of the energy 
density of our university is due to dark matter (DM). The last of the above observations points towards DM in the 
form of massive elementary particles, very likely having only (super)weak interactions with standard model particles ~\cite{Markevitch:2003at}. 

Though extra-terrestrial signals attributed to DM annihilation create sensations from time to 
time~\cite{Boehm:2003bt,Adriani:2008zr,Hooper:2010mq,Accardo:2014lma,Bulbul:2014sua}, alternative (and often more 
convincing) explanations in terms of astrophysical phenomena stand in the way of these becoming conclusive 
evidence~\cite{Hooper:2008kg,Boudaud:2014dta}. Under such circumstances, terrestrial verification of the 
existence of DM, in either direct searches for elastic scattering on nucleons~\cite{Undagoitia:2015gya} or 
in collider experiments~\cite{Abercrombie:2015wmb,Penning:2017tmb,Buchmueller:2017qhf}, is a desideratum.

Direct search experiments are likely to yield positive results if the DM consists of weakly interacting massive 
particles (WIMP)~\cite{Akerib:2018lyp,Aprile:2017iyp}. Such detection is unlikely for feebly interacting massive 
particles (FIMP)~\footnote{Note however that some future experiments will be able to probe typical FIMP-electron scattering 
cross-sections~\cite{Essig:2011nj,Alexander:2016aln,Crisler:2018gci}.}. Some scenarios not amenable to detection 
in direct searches can still lead to signals at colliders as much as those with WIMP DM. A well-known example is 
that  of supersymmetry where missing transverse energy (MET) at the Large Hadron Collider (LHC) is as likely a
signal for a neutralino DM in the minimal supersymmetric standard model (MSSM) as for the gravitino DM in 
gauge-mediated supersymmetry breaking scenarios~\cite{Feng:1997zr} with the associated visible particles serving 
as discriminator between the two cases~\cite{Bobrovskyi:2011vx,Brandenburg:2005he}
\footnote{Right-handed sneutrinos in certain simplified extensions of the MSSM 
can behave as WIMP DM candidates, which leave their footprints in the 
form of MET, in colliders~\cite{Belanger:2011ny,Dumont:2012ee,Banerjee:2013fga,Arina:2015uea}. Similar signatures are also 
obtained in supersymmetric $B-L$ extensions of the SM~\cite{Abdallah:2018gjj}.}. 
Interestingly, one can 
still envision other situations where direct searches are inconsequential on the one hand, while LHC signals,
on the other, are of a drastically different kind. A case in point is the MSSM augmented by right-chiral neutrino 
superfields, where a right-sneutrino becomes the DM candidate~\cite{Asaka:2005cn,Asaka:2006fs}~\footnote{It 
is important to note that a left-handed tau sneutrino, even when lighter than the 
lightest neutralino, will not serve as a thermal DM candidate, as it is excluded
by direct detection experiments~\cite{Falk:1994es,Arina:2007tm,Tan:2016zwf,Akerib:2016vxi,Chala:2017jgg}.}.

The above possibility is a  natural extension of the MSSM. Consider, as the simplest example, a right-chiral 
neutrino superfield  for each family, with just Dirac masses for neutrinos. Such a  superfield, being an SM 
gauge singlet, has only Yukawa interactions with the rest in the extended MSSM spectrum. Recent neutrino data 
constrain such couplings to rather small values ($y_{\nu}\,\simeq\,10^{-13}$)~\cite{Capozzi:2018ubv,Ade:2015xua}. 
If the sfermion masses evolve down to the TeV-scale from some high-energy values (not necessarily unified), then 
the mass parameters for all gauge non-singlet fields tend to go up through running induced by renormalisation 
group equations~\cite{Martin:1997ns}. Running of the mass parameter corresponding to $\snu$, the superpartners of 
right-handed neutrinos, is, however, negligibly small. Thus one of the right-sneutrinos is very likely to become 
the lightest supersymmetric particle (LSP) and  consequently a DM candidate in such a case. Moreover, the right 
stau ($\stau$) can quite conceivably become the next-to-lightest supersymmetric particle (NLSP)~\footnote{In 
fact the second lightest sneutrino, which we will assume to be almost degenerate with the sneutrino LSP, is 
strictly speaking  the NLSP. However, since the two additional sneutrinos have no impact 
on the collider phenomenology, we will loosely use NLSP to designate the lightest charged 
particle.}, since its Yukawa coupling is relatively large. The $\snu$, however, has extremely weak interactions with the rest of the MSSM 
spectrum, thus it typically does not reach thermal equilibrium with other particles in the early Universe.

As has been pointed out in a series of studies, such a scenario leads to a very characteristic signal in collider 
detectors if the NLSP is indeed the right-chiral stau~\cite{Banerjee:2016uyt,Evans:2016zau,Heisig:2011dr}. All SUSY cascades
at the LHC should then end up producing stau ($\stau$) pairs along with some SM particles. These stau 
($\stau$)-pairs will not decay into $\snu$s within the detector due to the small $y_{\nu}$ and will travel all 
the way through, leaving their signature as massive charged tracks. Such tracks can be distinguished from muonic 
tracks through event selection criteria such as track-$p_T$ and the time delay between the inner tracker and the 
muon chamber~\cite{Hinchliffe:1998ys}.

Since the signal and the SUSY spectrum here are both quite different from the well-studied case of a neutralino 
LSP, it is important to  reconstruct the superparticle masses in a scenario of this kind. Apart from collider 
phenomenology, the knowledge about the spectrum can reveal clues on the SUSY-breaking mechanism that is operative 
here. The $\snu$ DM candidate, of course, is illusive, since it is not even produced within the detector. The 
mass reconstruction procedures for neutralinos, charginos and left-chiral sleptons have been worked out in earlier
works~\cite{Biswas:2009zp,Biswas:2009rba,Chatterjee:2016rjo,Biswas:2010cd}. While the $\stau$-mass can be obtained from 
time-delay measurements, we pay special attention here to the mass reconstruction for the right-chiral smuon as 
well as the corresponding selectron, which thus yields a picture on the slepton flavour structure of the 
underlying theory.

In addition to the kinematic variables used earlier~\cite{Biswas:2009zp,Biswas:2009rba,Chatterjee:2016rjo,Banerjee:2016uyt,Biswas:2010cd,Gupta:2007ui}, 
notably the $p_{T}$ of the hardest jet and missing energy, $\met$, we have formulated event selection criteria based on additional quantities such as  the stransverse mass, $M_{T2}$~\cite{Lester:1999tx,Barr:2003rg}, to gain some insight into the 
right slepton  mass hierarchy. Our reconstruction procedure is applicable to right-smuons as well as selectrons 
for both the cases where they are heavier and lighter than the lightest neutralino ($\neu$). 

The paper is organised as follows: In section~\ref{sec:model} we discuss the model considered along with the 
constraints imposed from both colliders results and cosmology. In section~\ref{sec:strategy} we discuss the 
supersymmetric signals that we analyse, along with the strategy for the reconstruction of the slepton masses.
Section~\ref{sec:results} contains the benchmark points chosen for different case studies together with an 
analysis of the discovery prospects corresponding to the signatures considered in upcoming runs of LHC at an 
integrated luminosity of $\mathcal{L}\,=\,3000$ fb$^{-1}$. The $M_{T2}$ and slepton mass ($m_{\tilde{l}}$) 
distributions for the two different mass orderings considered are also studied in section~\ref{sec:results}. 
Finally we summarise and conclude in section~\ref{sec:summary}.

\section{The theoretical scenario, the spectrum and its constraints}\label{sec:model}
We consider the MSSM supplemented with three families of right-handed (RH) neutrino superfields ($\hat{\nu}_{R}$) 
with Dirac mass terms for the neutrinos. Hence the superpotential (suppressing family indices) becomes 
\begin{equation}
W = W_{MSSM} + y_{\nu}\,\hat{H}_{u}\,\hat{L}\hat{\nu}^{c}_{R},
\label{suppot}
\end{equation}  
where $W_{MSSM}$ is the superpotential of the MSSM, $y_{\nu}$ is the neutrino Yukawa coupling, $\hat{L} = 
(\hat{\nu}_{L}, \hat{l}_{L})$ is the left-handed (LH) lepton superfield and $H_u$ is the Higgs doublet that 
couples to the up-type quarks. The physical states dominated by right sneutrinos ($\snu$) have all their 
interactions proportional to $y_{\nu}$. For simplicity, we consider a scenario where all (right) sneutrinos are
degenerate and the sneutrino mass matrix is diagonal. After electroweak symmetry breaking, the neutrinos acquire 
masses as shown below
\begin{equation}
m_{\nu} = \frac{y_{\nu}}{\sqrt{2}}\,v\,\sin\beta,
\label{neumass}
\end{equation}
where $v \simeq 246.2$ GeV and $\tan\beta = \frac{\langle\,H_{u}^{0}\rangle}{\langle\,H_{d}^{0}\rangle}$. Recent
data from global fits on neutrino oscillation and cosmological bound on the sum of neutrino masses, constrain 
the largest Yukawa coupling in the range $2.8\times10^{-13}\,\lesssim\,(y_{\nu}\,\sin\beta)\,\lesssim\,
4.4\times10^{-13}$~\cite{Banerjee:2016uyt}. The lower bound is taken from a global fit on the neutrino 
oscillation parameters in the normal hierarchy scenario ~\cite{Capozzi:2018ubv}, while the upper bound is 
obtained from a combination of Planck, lensing and baryon acoustic oscillation data~\cite{Ade:2015xua}. The 
latter bound can vary roughly by a factor of two depending on the set of cosmological data included in the 
fit~\footnote{For a recent compilation see Ref.~\cite{Lattanzi:2017ubx}.}.

Barring the right neutrino superfields, we consider the phenomenologically constructed MSSM 
(pMSSM)~\cite{Djouadi:1998di}. Thus the soft SUSY breaking terms are free parameters. The addition of the RH 
neutrino superfield entails the following additional soft terms in the MSSM Lagrangian:
\begin{equation}
-\mathcal{L}_{soft}\,\supset\,m_{\snu}^{2}|\snu|^{2}+(y_{\nu}\,A_{\nu}\,H_{u}\,\tilde{L}\,\snu^{c} + h.c.),
\label{softterms}
\end{equation}
where $A_{\nu}$ plays a role in the left-right mixing in the sneutrino sector. The sneutrino mass matrix is
defined as
\begin{center}
\begin{equation}
{\cal M}_{\tilde\nu}^2 =
\begin{bmatrix}
m_{\tilde{\nu}_L}^2 &  -y_{\nu}\,v\,\sin\beta\,(\mu\cot\beta\,-\,A_{\nu}) \\
-y^{\dagger}_{\nu}\,v\,\sin\beta\,(\mu^{*}\cot\beta\,-\,A^{*}_{\nu}) &  m_{\snu}^2 \\
\end{bmatrix}
\end{equation}
\end{center}
where $m_{\tilde{\nu}_L}$ and $m_{\snu}$ are respectively the soft scalar masses for the left- and right-chiral 
sneutrinos. One then finds that the left-right sneutrino mixing angle, $\tilde{\Theta}$, can be written as 
\begin{equation}
\tan\,2\tilde{\Theta} = \frac{2\,y_{\nu}\,v\,\sin\beta\,|\mu\cot\beta\,-\,A_{\nu}|}{m_{\tilde{\nu}_{L}}^2 - m_{\snu}^2},
\label{mixangle}
\end{equation}
thus implying that the admixture of $SU(2)$ doublets in the $\snu$-dominated mass eigenstates are limited by the 
neutrino Yukawa couplings. 

As mentioned in the introduction, the present study focuses on scenarios with the lighter stau ($\staul$) as 
the NLSP. 
Such a stau, upon production at the LHC, will eventually decay into the right sneutrino LSP through modes such 
as $\tilde{\tau_{1}}\,\rightarrow\,W^{(*)}\snu$, driven, as expected, by the neutrino Yukawa coupling. For 
$m_{\tilde{\tau_{1}}}\,>\,m_{\snu}\,+\,m_{W}$, the width of the above two-body decay is given by
\begin{equation}
\Gamma_{\tilde{\tau_{1}}}\,\simeq\,\Gamma_{\tilde{\tau_{1}}\,\rightarrow\,W\snu}\,=\,\frac{g^2\tilde{\Theta}^2}{32\,\pi}|U_{L1}^{(\tilde{\tau_{1}})}|^2\,\frac{m_{\tilde{\tau_{1}}}^3}{m_{W}^2}\,\left[1-\frac{2(m_{\snu}^2+m_{W}^2)}{m_{\tilde{\tau_{1}}}^2}\,+\,\frac{(m_{\snu}^2 - m_{W}^2)^2}{m_{\tilde{\tau_{1}}}^4}\,\right]^{3/2},
\label{decaywidth} 
\end{equation}
where $g$ is the $SU(2)_{L}$ gauge coupling, $m_{W}$ the $W$-boson mass and $U^{(\tilde{\tau_{1}})}$ 
parametrises the left-right mixing of the staus. Assuming $A_{\nu}$ is of the same order as the other trilinear 
couplings, the $\staul$s are fairly long-lived with a typical life-time of $\mathcal{O}(1)\,$sec for all the 
benchmark points that we will consider in section~\ref{sec:results}. Thus, the decay length of $\tilde{\tau_{1}}$ 
is large compared to the typical collider scale. All processes at the LHC, which are initiated with the 
production of superparticles, will ultimately lead to the production of a pair of quasi-stable $\tilde{\tau_{1}}$s 
which will travel all the way up to the muon-chamber. In addition to making the NLSP stable at the collider 
scale, the smallness of $y_{\nu}$ also implies an out-of-equilibrium decay of the NLSP in the early universe
into the $\snu$ LSP. 
The contribution to the $\snu$ relic density has two components. The 
first of which arises from the decay of the stau after it freezes out,
and can be estimated as
\begin{equation}
\Omega^{FO}_{\snu}h^{2} = \frac{m_{\snu}}{m_{\staul}}\,\Omega_{\staul}h^{2},
\label{eq:FOrelic}
\end{equation}  
where $\Omega_{\staul}h^{2}$ is the (thermal) relic density of the quasi-
stable NLSP when it freezes out. This contribution can be calculated 
via $\Omega_{\staul}h^{2}$ using a standard package such as
$\texttt{microOMEGAs}$~\cite{Belanger:2014vza}.

In addition to the contribution ensuing from the out-of equilibrium decay of the $\staul$ NLSP, 
the remaining heavy supersymmetric particles, \textit{viz.}, the left-handed sleptons, left-handed 
sneutrinos, neutralinos, charginos etc., may also decay while still in thermal equilibrium. These latter contributions which arise from the freeze-in mechanism~\cite{Hall:2009bx,McDonald:2001vt} can be approximated as
\begin{equation}
\Omega^{FI}_{\snu}h^{2}\,\simeq\,\frac{1.09\times10^{27}}{g^{*\,3/2}}m_{\snu}\underset{i}{\sum}\,\frac{g_{i}\Gamma_{i}}{m^{2}_{i}}
\label{eq:FIrelic}
\end{equation}
where $g^{*}\,\approx \,106.75$~\cite{Asaka:2005cn}, is the average number of 
effective degrees of freedom contributing to the thermal bath,
and the sum runs over all the aforementioned relevant 
superparticles. Besides, $\Gamma_{i}$, $m_{i}$ and $g_{i}$ are respectively the decay width to $\snu$, 
mass, and degrees of freedom of the $i^{th}$ superparticle. The decay widths of several such superparticles 
into $\snu$ are listed in Ref.~\cite{Asaka:2005cn}. Thus, the total relic density of the sneutrinos is 
given as 
\begin{equation}
\Omega_{\snu}h^{2}\,=\Omega^{FO}_{\snu}h^{2}+\Omega^{FI}_{\snu}h^{2}
\label{eq:relic}
\end{equation}

We by and large assume the three right-handed sneutrinos to be mass 
degenerate. However, this assumption may not be realised in practice, and 
one may encounter 
small splittings among the three families. In such cases, the heavier right-handed mass eigenstates, 
\textit{viz.}, $\snu^{e\,,\mu}$, may in principle be produced from the decay of heavier superparticles 
following equation~\eqref{eq:FIrelic}. These $\snu^{e\,,\mu}$ when produced, will ultimately decay into 
the $\snu$ LSP. However, these decays are suppressed by two powers of the neutrino Yukawa coupling, and 
hence almost always have lifetimes greater than the present age of the universe. Therefore, the two other 
$\snu$-dominated states will make a substantial contribution to the relic density regardless of whether the
the three $\snu$s are mass degenerate or not. Thus, the $\Omega^{FI}_{\snu}h^{2}$ must also include the 
abundances of $\snu^{e\,,\mu}$.

So far, we have discussed only about the $\snu$ LSP and $\staul$ NLSP. However, depending on the details of 
the SUSY breaking scheme, one can have various mass hierarchies in the non-strongly interacting superparticle 
sector, particularly in the masses of the right-chiral smuon and the selectron, which we assume to be 
degenerate and heavier than the stau NLSP, with respect to the lightest neutralino mass, $m_{\neu}$. Hence, 
one may encounter two distinct mass orderings between these particles, \textit{viz.},
\be
\; {\rm Case} \;\;  {\rm I} \;\;:\;\; m_{\neu} > m_{\smu} = m_{\sel},
\label{caseA}
\ee
and
\be
\; {\rm Case } \;\; {\rm II} \;\;:\;\;  m_{\smu} = m_{\sel} > m_{\neu},
\label{caseB}
\ee
These different hierarchies may leave their markedly unique footprints in collider signals. Hence, 
experimentally identifying the relevant mass ordering may unveil the physics behind the SUSY breaking.
Thus, the main focus of this present work is to understand the effects of these hierarchies on the collider
signals and to devise strategies to separate one from the other. However, before detailing the analyses
dedicated solely for the discrimination in the two hierarchies at the high luminosity run of the LHC (HL-LHC), 
we ensure that our benchmark points satisfy all the following constraints.
\begin{itemize}
\item The mass of the lightest CP-even Higgs is required to lie in the range $123 \; \textrm{GeV} \; < m_{h^{0}} < 128$ GeV, 
which is consistent with the Higgs mass measurements from various channels at the LHC~\cite{Aaboud:2018wps,Aad:2015zhl}. 

\item
The signal strengths of the SM-like Higgs boson are required to lie within the experimentally measured values
and their uncertainties~\cite{Khachatryan:2014jba,Aad:2015gba}. We use $\texttt{LILITH}$~\cite{Bernon:2015hsa} in order to compute the 
likelihood function and require them to be $\sim 1$ for all our chosen benchmark points (BPs).
Furthermore, we also perform a cross-check and find that the signal strengths in the individual 
Higgs decay channels lie within their experimental uncertainties upon employing the 
\texttt{HiggsBounds} package~\cite{Bechtle:2013wla}.

\item We impose that the relic density of the LSP, $\Omega_{\snu}h^{2}$, satisfies the  upper bound (at the 2$\sigma$ level) obtained 
by  PLANCK, namely $\Omega_{DM}h^{2}\,=\,0.1199\pm0.0027$~\cite{Ade:2015xua}.  

\item A long-lived charged particle with  hadronic decay modes can affect the standard Big Bang Nucleosynthesis 
(BBN). In particular, it may lead to an over-prediction of the abundance of light nuclei like deuterium. In 
order to avoid destroying the successful predictions of the light element abundance, we require that the stau 
NLSP lifetime does not exceed 100 seconds~\cite{Kawasaki:2004qu,Banerjee:2016uyt,Kawasaki:2017bqm}.

\item The current model-independent studies on heavy stable charged tracks from the LHC requires $\staul\,>$ 360 
GeV, as obtained by CMS for a pair produced scenario~\cite{CMS-PAS-EXO-16-036}.

\item Furthermore, we demand the gluino and squark masses to be $m_{\tilde{g}}\,>\,2.1$ TeV, $m_{\tilde{q}}\,>\,1.4$ 
TeV and $m_{\tilde{t}}\,>\,1.1$ TeV from recent available bounds from the LHC~\cite{Aaboud:2017vwy,Sirunyan:2017kqq}.
These limits are based on searches in the jets + missing energy channel, which are relevant for the MSSM with a 
neutralino LSP. However, in the absence of any dedicated SUSY search results based on stable charged track 
signals, we conservatively use the aforementioned limits.
\end{itemize}

\section{Mass reconstruction  strategy}\label{sec:strategy}
In order to decipher the actual ordering of the masses in the SUSY electroweak sector, in particular of $\neu$ 
and $\smu/\sel$, we have to reconstruct the following three particles, \textit{viz.}, $\staul$, $\neu$ and 
$\smu/\sel$, with the $\smu$ and $\sel$ being considered to be degenerate in mass. As discussed above, the mass 
of the $\staul$ can be  reconstructed using the time-of-flight measurements following~\cite{Banerjee:2016uyt,Hinchliffe:1998ys} 
while the  neutralino($\neu$) reconstruction can easily be performed using the procedure envisioned in 
Ref.~\cite{Biswas:2009zp}. For completeness, we briefly summarise these two strategies.

\noindent
{\bf $\staul$ reconstruction:}
As $\staul$s are very heavy, typically ${\cal O}(100)~\rm{GeV}$ particles, they are slow. Their velocity 
distributions can be obtained using the time delay between the production of $\staul$s at the interaction point 
and their detection in the muon chamber. Combining this with the momentum measured in the muon chamber, one can 
reconstruct the $\staul$ mass by exploiting the relation,

\begin{equation}
m_{\staul}\,=\,\frac{p}{\beta\,\gamma}\,,
\label{eqn:staumass}
\end{equation}

\noindent
where $p,\beta$ and $\gamma$ are respectively the momentum, speed with respect to the speed of light and the 
Lorentz factor, of the $\staul$. In order to be fairly realistic with the experimental situation, we smear the 
actual velocity of $\staul$s with the Gaussian (Box-Muller) prescription by choosing a standard deviation of 
$\sigma_{\beta}\,=\,0.024$, upon following ATLAS' calibration as shown in Fig. 1 (right) in Ref.~\cite{ATLAS:2014fka}. 

\noindent
{\bf $\neu$ reconstruction:}
In order to reconstruct the $\neu$ mass, one may look for 2$\staul$ + 2$\tau$ states, dominantly produced by 
$\tilde{q},\tilde{g}$ initiated cascades. The invariant mass distributions of these $\staul+\tau$ pairs will 
peak around the $\neu$ mass. The most challenging part of this technique is the reconstruction of $\tau$s 
because of their semi-invisible decays. To tackle this difficulty of reconstructing the $\tau$ masses, we employ 
the collinear approximation as described below.

\noindent
{\bf Collinear Approximation:}
Following the method described in~\cite{Rainwater:1998kj}, one can fully reconstruct the $\tau$s with the 
knowledge of the fraction, $x_{\tau_{c_{i}}}$ ($i=1,2$), of the parent $\tau$ momentum carried by the ensuing 
visible charged jet or lepton. Each event has two unknowns, \textit{viz.}, the two components of the momenta of the neutrinos (one 
(two) neutrino(s) per hadronic (leptonic) $\tau$ decay). These two unknowns can be solved on an event-by-event 
basis upon knowing the two components of the missing transverse energy, $\vec{\slashed{E}}_T$. If 
$p_{\tau_{i}}^{\mu}$ and $p_{\tau_{c_{i}}}^{\mu}$ are the four momenta of the two parent $\tau$-leptons and the 
corresponding visible charged objects, then one has
\be
p_{\tau_{c_{i}}}^{\mu} = x_{\tau_{c_{i}}}\,p_{\tau_{i}}^{\mu},
\ee
and one obtains
\be
\vec{\slashed{E}}_T = \Big(\frac{1}{x_{\tau_{c_{1}}}}\,-\,1\Big)\vec{p}_{\tau_{c_{1}}}\,+\,\Big(\frac{1}{x_{\tau_{c_{2}}}}\,-\,1\Big)\vec{p}_{\tau_{c_{2}}},
\ee 
where the $\tau$ has been considered to be massless and the neutrinos from these $\tau$ decays are assumed to
be collinear in the direction of their corresponding visible charged objects. Provided the decay products are 
not back-to-back, the above equation provides two conditions (from the $x$- and $y$-components of 
$\vec{\slashed{E}}_T$) for $x_{\tau_{c_{i}}}$ and one finally obtains the $\tau$ momenta as 
$p_{\tau_{c_{i}}}/x_{\tau_{c_{i}}}$. 

\begin{figure}[t]
\begin{center}
\includegraphics[width=5.0cm]{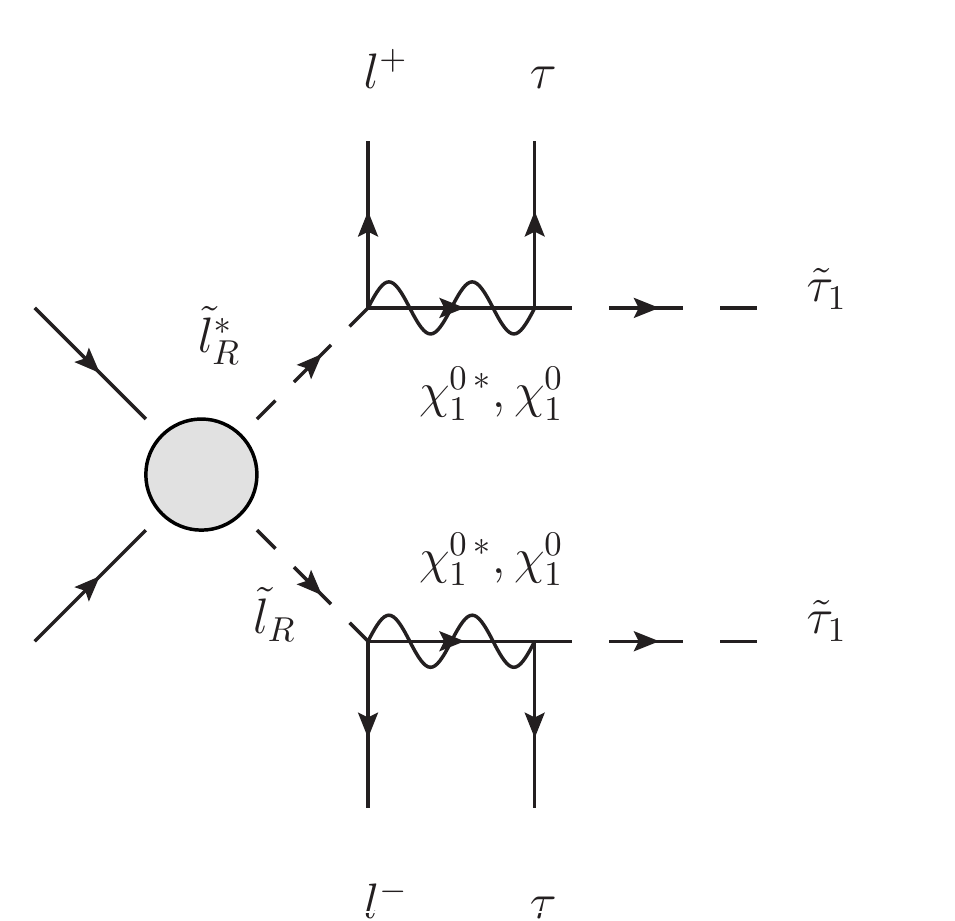}
\includegraphics[width=8.0cm]{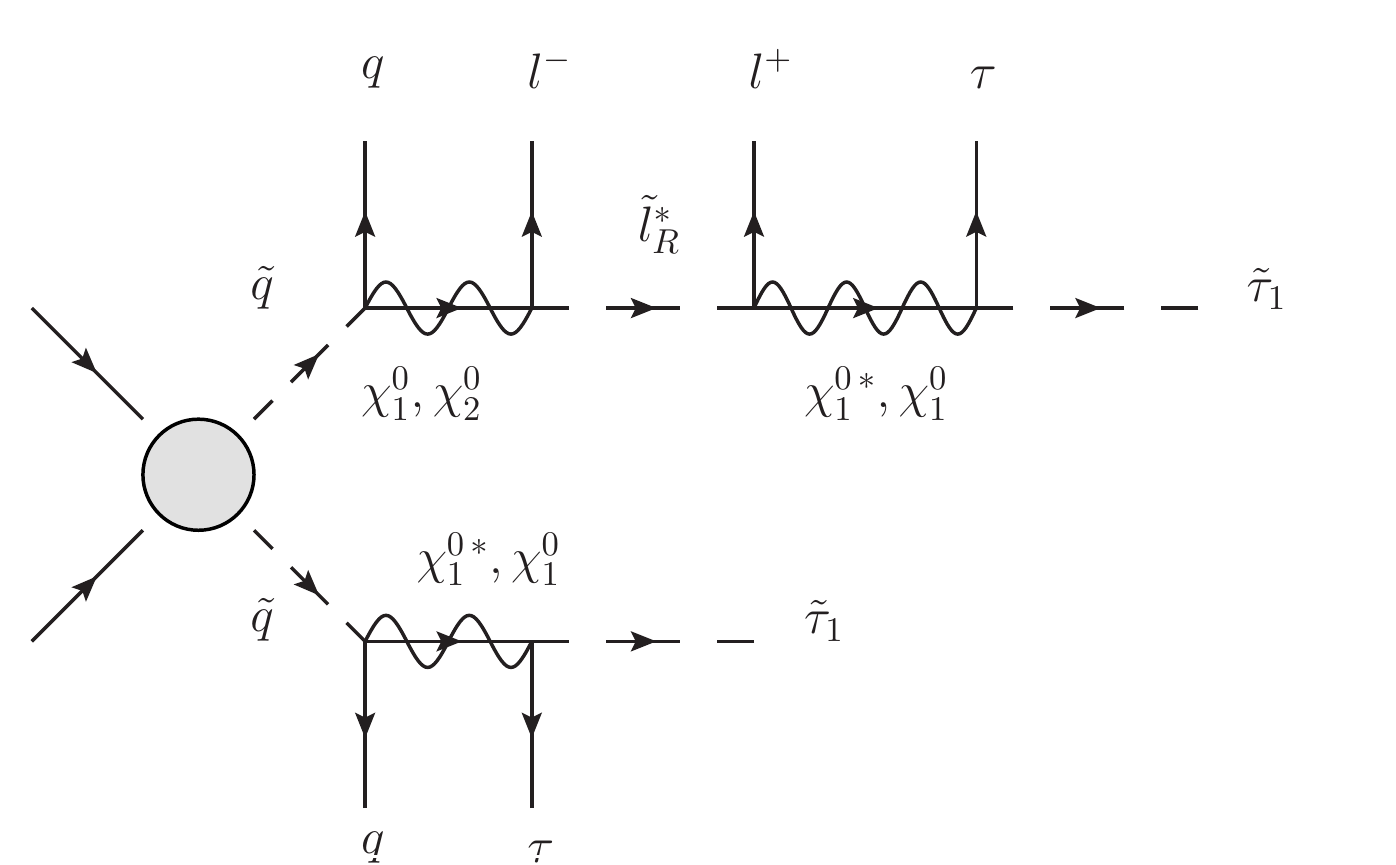}
\caption{Representative diagrams for the Drell-Yan production of $\tilde{l}_{R}$ is shown in the left. The 
right panel illustrates  a SUSY cascade process  initiated by SUSY particles from the strong sector that mimics 
the final state of the left panel, modulo hard jets.}
\label{fig:signal}
\end{center}
\end{figure}

\noindent
{\bf Slepton reconstruction:}
Finally, in order to reconstruct the slepton masses ($m_{\smu}, m_{\sel}$), we consider the  Drell-Yan production 
of $\,\smu\,\smu^{*}\,(\sel\,\sel^{*})$ followed by the slepton's decay into a lepton ($\mu/e$), a $\staul$ and 
a $\tau$, mediated by an off-shell or an on-shell $\neu$ depending on their mass ordering. The topology of the 
process is shown in the left panel of Fig.~\ref{fig:signal}. In the following, for both hierarchies mentioned in Eqs.~\ref{caseA} and~\ref{caseB}
we investigate the two possible  signatures,

\begin{itemize}
\item{} 2 $\staul$s + 2 opposite sign same flavour leptons (OSSF) + 1 $\tau$-tagged jet + $\met$ 
\item{} 2 $\staul$s + 2 opposite sign same flavour leptons (OSSF) + 2 $\tau$-tagged jet + $\met$.
\end{itemize}
 
To reconstruct the slepton mass, we utilise the popular stransverse mass variable, $M_{T2}$~\cite{Lester:1999tx,Barr:2003rg}. 
In general, $M_{T2}$ is a useful variable for measuring the mass of a particle when it is pair-produced in a 
hadron collider and thereafter decays into a visible object along with invisible particles, thus giving rise to 
missing transverse momentum. Hence, the $M_{T2}$ variable can be relevant for the reconstruction of 
slepton masses for the first signature involving a single $\tau$-tagged jet. The variable $M_{T2}$ is defined as
\begin{equation}
M_{T2}\,\equiv\,\underset{\vec{\slashed{p}}_{T,1}+\vec{\slashed{p}}_{T,2}=\vec{\slashed{E}}_T}{\text{min}}\left(\,\text{max}\,\{m_{T}(\vec{p}_{T,1},\vec{\slashed{p}}_{T,1},m,m_{inv})\,,m_{T}(\vec{p}_{T,2},\vec{\slashed{p}}_{T,2},m,m_{inv})\}\right) , 
\label{eqn:symMT2} 
\end{equation}
where $m, \vec{p}_{T,i}, \vec{\slashed{p}}_{T,i}$ and $m_{inv}$ are respectively the mass of the visible objects, 
transverse momenta of the visible objects, the missing transverse momenta and the mass of the invisible particles 
in the $i^{th}$ leg and $m_T$ refers to the standard transverse mass variable. The actual mass of the mother 
particle will always be bounded from below by $M_{T2}$ and hence the end point of the $M_{T2}$ distribution will 
give a fairly accurate estimate of its mass. The above definition is slightly modified to accept asymmetries, 
which leads us to the asymmetric $M_{T2}$ variable~\cite{Konar:2009qr} and is shown to be more useful than its 
symmetric counterpart while reconstructing the slepton masses. The asymmetric $M_{T2}$ variable is defined as
\begin{equation}
M_{T2}\,\equiv\,\underset{\vec{\slashed{p}}_{T,1}+\vec{\slashed{p}}_{T,2}=\vec{\slashed{E}}_T}{\text{min}}\left(\,\text{max}\,\{m_{T}(\vec{p}_{T,1},\vec{\slashed{p}}_{T,1},m_{1},m_{inv,1})\,,m_{T}(\vec{p}_{T,2},\vec{\slashed{p}}_{T,2},m_{2},m_{inv,2})\}\right) , 
\label{eqn:asymMT2} 
\end{equation}
with different masses for the invisible particles in the two legs. For our first scenario involving a single $\tau$-tagged jet, the other visible $\tau$ decay product
escape the detector undetected. Hence the two visible particles (in each leg) required to construct the asymmetric $M_{T2}$ are the $\staul$ along 
with its nearest (in the $\eta$-$\phi$ plane) lepton. The $\tau$-tagged jet is considered to be part of that leg 
for which it is nearest (in its $\Delta R=\sqrt{\Delta \eta^2 + \Delta \phi^2}$ separation) to the $\staul$ and 
it is thus combined with the corresponding $\staul \ell$ pair. Hence, both the $\nu_{\tau}$ and the undetected 
$\tau$-jet/$\tau$-lepton contribute to $\met$ in this case. Owing to the smallness of the mass of the 
$\tau$, one can safely use $ m_{inv,1}\,=\,m_{inv,2}\,\simeq\,0$ while constructing the $M_{T2}$ variable. 
The asymmetric $M_{T2}$ variable constructed in this way will be bounded above by $m_{\smu(\sel)}$.

For the signature involving double $\tau$-tagging, we fully reconstruct both the sleptons upon 
using the invariant masses of the three individually reconstructible objects, \textit{viz.}, $\staul,\tau 
\; \textrm{and} \;\ell^{\pm}$. The $\tau$s for this analysis are reconstructed according to the collinear approximation discussed
above. In order to reconstruct both the sleptons properly, we construct all possible pairs of invariant masses
$m_{\staul\,\tau\,l^{\pm}}$ and compute the difference between the invariant masses of each pair. The pair yielding the least
difference in the invariant mass is regarded as the correct pair. 
  
Lastly, pair production of strongly interacting superparticles also leads to  similar final states but 
exhibit different topologies (as shown in Fig~\ref{fig:signal} (right)), namely the 
cascade decay has additional jets at the parton level. Hence, one cannot use the 
aforementioned procedures for slepton mass reconstruction for such processes. 
Our strategy is to choose cuts in order to suppress the contribution of processes initiated by strongly interacting particles. 
The cascade  processes will always give rise to harder jets, to higher jet multiplicities and to a harder $\met$ distribution as compared with DY production. Hence 
a  hard cut on the $p_{T}$ of the hardest jet as well as a cut on the jet multiplicity for jets above 
a certain threshold $p_{T}$ $\sim100$ GeV, and a hard upper cut on the $\met$ can efficiently reduce the effects of the cascade processes, as we will see
below. Moreover as $\met$ plays  an important role in the construction of the $M_{T2}$ variable, which will, at the end be our most 
important observable for the mass reconstruction of the sleptons, removing the cascade processes with this cut 
will help in achieving faithful  reconstructions of the sleptons.

\subsection{Signal and background}

In the remaining part of this section, we focus on the various details of our collider analyses. The 
presence of $\staul$s in the signal makes it easier to reduce the major SM backgrounds ensuing from 
the two real backgrounds, \textit{viz.}, $ZZZ$ and $Zh$ and the following fakes, $ZZ$, $t\bar{t}Z$, 
and $ZW^{+}W^{-}$. All these SM backgrounds have been merged with up to two additional partons upon 
employing the MLM merging scheme~\cite{Mangano:2006rw} with appropriate choices for merging parameters. 
We ensure at least two muons, exactly two taus and two additional leptons (electrons or muons) for 
the real backgrounds. For the fake backgrounds, the additional merged jets will fake the tau jets or 
the leptons, as we will discuss below. As has already been discussed, the long-lived $\staul$s have 
signatures similar to muons, but are much heavier. Because of their significantly large mass, these 
particles are sluggish, having much lower velocities than their SM muon counterparts. The $p_T$ and 
velocity distributions of the $\staul$s will be utilised to discriminate them from the SM muons. 
Following the footsteps of certain experimental analyses~\cite{Chatrchyan:2013oca,ATLAS:2014fka}, we 
impose a hard cuts on the $p_{T}$ of the two hardest muons (or $\staul$s for the signal) with an 
additional requirement of the $\staul/\mu$ speed to be $\beta(=\frac{p}{E})\,<\,0.95$.

For the collider analyses, we generate the signal and background samples along with their decays in 
the $\texttt{MadGraph5\_aMC@NLO}$~\cite{Alwall:2014hca} framework. The parton showering and 
hadronisation is done in $\texttt{Pythia 8}$~\cite{Sjostrand:2014zea}. The jets are constructed with 
the anti-$kT$~\cite{Cacciari:2008gp} algorithm with a minimum $p_T$ of 20 GeV and a jet parameter of 
$R = 0.4$, using the \texttt{FastJet}~\cite{Cacciari:2011ma} package. Finally, we perform a fast 
detector analysis in the $\texttt{Delphes 3}$ framework~\cite{deFavereau:2013fsa}. For all sample 
generations, we use the \texttt{NNPDF2.3}~\cite{Ball:2012cx} parton distribution function set, at
leading order (LO). The renormalisation and factorisation scales are set to the default dynamic values 
in $\texttt{MadGraph5\_aMC@NLO}$. For the signal samples however, we use flat $K$-factors to 
approximately capture the next to leading order (NLO) effects. For this purpose, we determine our 
signal cross-sections at NLO with $\texttt{Prospino2.0}$~\cite{Beenakker:1996ed} and scale the LO 
samples accordingly. Flat NLO $K$-factors for the backgrounds are computed within 
$\texttt{MadGraph5\_aMC@NLO}$ by taking the ratios of the unmerged cross-sections at NLO and LO. We
scale the merged $ZZZ, Zh, ZZ, t\bar{t}Z$ and $ZW^+W^-$ cross-sections by 1.53, 2.17 (which also 
includes a correction factor to the Higgs branching ratio), 1.48, 1.32 and 2.03 respectively. For 
the detector-level analyses, we employ the following cuts:

\begin{itemize}
\item For the two hardest muons ($\staul$s in the case of our signal), we require the transverse
momenta of the these two objects to be $p_{T}^{\mu_{1,2}}>70$ GeV, the speed, $\beta^{\mu_{1,2}}< 0.95$
and the rapidity to lie in the range, $|\eta(\mu_{1,2})| < 2.5$. Furthermore, we require these objects
to be separated in the $\eta$-$\phi$ plane by $\Delta\,R(\mu_{1},\mu_{2}) > 0.4$.

\item For the remaining opposite-sign-same-flavour (OSSF) leptons ($e,\mu$), we require, $p_{T}^{\ell} 
> 10$ GeV, $\beta(\ell) > 0.95$, $|\eta(\ell)| < 2.5$ and $\Delta R(\ell_{1},\ell_{2}) > 0.2$.

\item For all jets (quark/gluon initiated as well as $\tau$-tagged ones), we demand the jets to have
$p_{T}^{j} > 20$ GeV, $|\eta(j)| < 5$ and $\Delta R(j,j) > 0.4$.

\item In addition, we require, $\Delta R(\mu_{1,2},j) > 0.4$ and $\bigtriangleup\,R(\ell,j) > 0.4$.
\end{itemize} 

\begin{table}[t]
\begin{center}
\begin{tabular}{|c|c|c|c|}
\hline
Parameter  & $p_{T}(j_{1})$ & Number of jets with $p_{T}(j) > 100$GeV & $\met$       \\\hline\hline
Cut set A  & $< 200$ GeV    & $< 2$                                   & $< 150$ GeV  \\
Cut set B  & $< 200$ GeV    & $< 2$                                   & $< 200$ GeV  \\ \hline
\end{tabular}
\caption{Selection cuts applied to suppress the squark-gluino processes. Here $p_T(j_1)$ refers
to the transverse momentum of the hardest jet.}
\label{cut1}
\end{center}
\end{table}

\begin{figure}[t]
\begin{center}
\includegraphics[width=5.0cm, angle=270]{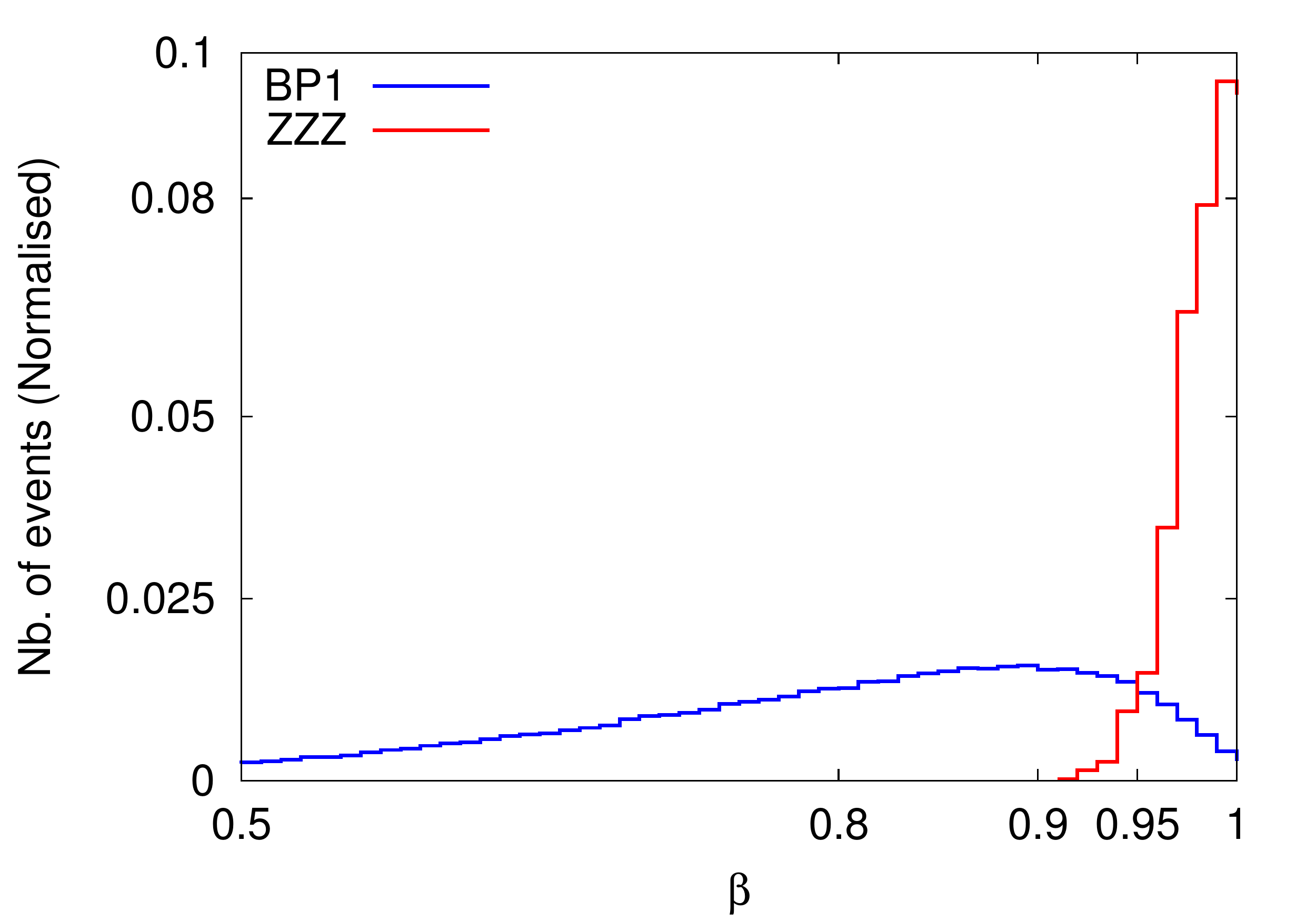}
\caption{$\beta$-distribution for signals as well as background events 
are shown. The distribution clearly suggests that a negligible number of 
background events survives after the application of the $\beta$-cut as 
mentioned in the text.}
\label{fig:beta_dist}
\end{center}
\end{figure}

Moreover, in order to suppress the squark-gluino contamination, we implement the additional cuts
 listed in Tab.~\ref{cut1}. In Fig.~\ref{fig:beta_dist}, we sketch the $\beta$-distribution of
the hardest muon/$\staul$ for BP1 (as defined below) and for the $ZZZ$ background, with the following 
values for the mean and rms, $\mu_{sig} = 0.768, \; \mu_{bkg} = 0.999$ and $\sigma_{sig} = 
0.167, \; \sigma_{bkg} = 0.024$~\footnote{The mean and the rms for the background are a result of the
Gaussian smearing introduced by hand.}. One can clearly see that requiring $\beta \lesssim 0.95$ strongly suppresses the SM background events. Thus, 
geared with this setup, we proceed with the reconstruction of the slepton masses in the following 
section.

\section{Results}\label{sec:results} 
In this section, we utilise the entire arsenal of techniques discussed above to finally show the 
viability of the slepton reconstructions and illustrate the possibility of probing the two mass 
hierarchies. For this purpose, we choose six benchmark points from the pMSSM spectrum, augmented 
with three additional families sneutrino fields. We ensure that all these BPs abide by the constraints listed in 
section~\ref{sec:model}. Three of these BPs correspond to the case $m_{\neu} > m_{\smu,\sel}$ and 
are summarised in Tab.~\ref{BPcaseA}. The remaining three corresponding to  $m_{\neu} < m_{\smu,\sel}$ are 
shown in Tab.~\ref{BPcaseB}.  

Before commencing the collider study, we make a small digression to 
explain the factors contributing to the relic density. As is evident from 
Tables~\ref{BPcaseA} and~\ref{BPcaseB}, for all our benchmark points, 
the $\snu$ relic density is in agreement with the value reported by the 
PLANCK collaboration~\cite{Ade:2015xua}. The dominant contribution comes
from the \textit{freeze-in} mechanism~\footnote{As an 
example, for BP3, with $A_{\nu}=-2619$ GeV and $m_{\snu}=39.2$ GeV, we obtain $\Omega^{FI}_{\snu}h^{2} \sim 
0.114$ and $\Omega^{FO}_{\snu}h^{2} \sim 0.006$. It might however be 
possible to have a  larger freeze-out fraction by increasing the mass of the 
decaying supersymmetric particle as in such case the \textit{freeze-in} contribution (Eq.~\eqref{eq:FIrelic}) is suppressed relative to the \textit{freeze-out} contribution (Eq.~\eqref{eq:FOrelic}).} Even though the mass 
of the sneutrino LSP is not relevant for the collider analysis that follows, 
it directly affects the the relic density as is evident from equations~\eqref{eq:FOrelic} and~\eqref{eq:FIrelic}. The neutrino 
trilinear coupling ($A_{\nu}$) is also a deciding factor since it determines the decay widths that 
control $\Omega^{FI}_{\snu}h^{2}$. On the one hand, large values of $A_{\nu}$ imply large 
$\Omega^{FI}_{\snu}h^{2}$. At the same time, a small $A_{\nu}$ increases the lifetime of $\staul$, 
thereby increasing the possibility of being strongly constrained by the BBN.  
As an example, in the case of BP3, as we increase $A_{\nu}$ from -2619 GeV 
to -400 GeV, the allowed value of $m_{\snu}$ increases from 39.2 GeV to 52.2 
GeV~\footnote{For this case, the $\Omega^{FI}_{\snu}h^{2}$ and 
$\Omega^{FO}_{\snu}h^{2}$ change to $\sim 0.112$ and $\sim 0.009$ 
respectively, still keeping the \textit{freeze-in} contribution almost an 
order of magnitude larger than its \textit{freeze-out} 
counterpart.}, while the $\staul$ lifetime increases from $\sim 2$ seconds 
to $\sim 94$ seconds. Therefore, in our analysis we have fixed $A_{\nu}$ 
around the TeV-scale and thereby determine the allowed sneutrino mass 
($m_{\snu}$) in order to saturate the abundance.

\begin{table}[t]
 \begin{center}
\begin{tabular}{|c|c|c|c|}
\hline
Masses (in GeV)                                  & BP1  & BP2   & BP3 \\\hline\hline
$m_{\tilde{g}}$                                & 2235 & 2200  & 2224   \\
 $m_{\tilde{u}_{L}}\,,m_{\tilde{c}_{L}}$       & 2004 & 2023  & 2124  \\ 
 $m_{\tilde{u}_{R}}\,,m_{\tilde{c}_{R}}$       & 1922 & 1919  & 2020  \\
 $m_{\tilde{d}_{L}}\,,m_{\tilde{s}_{L}}$       & 2005 & 2025  & 2125  \\
 $m_{\tilde{d}_{R}}\,,m_{\tilde{s}_{R}}$       & 1914 & 1920  & 2020  \\ 
 $m_{\tilde{t}_{1}}$                            & 1218 & 1266  & 1373  \\
 $m_{\tilde{t}_{2}}$                             & 1764 & 1741  & 1843  \\
 $m_{\tilde{b}_{1}}$                               & 1705 & 1692  & 1797  \\
 $m_{\tilde{b}_{2}}$                               & 1740 & 1732  & 1840  \\ 
 \hline \hline
 $m_{\chi^{0}_{2}}$                                  & 802  & 1009  & 942 \\
 $m_{\chi^{\pm}_{1}}$                  		       & 802  & 1009  & 913  \\
 $m_{\tilde{\nu}_{e_{L}}}\,,m_{\tilde{\nu}_{\mu_{L}}}$ & 896  & 901 & 1011\\
 $m_{\tilde{\nu}_{\tau_{L}}}$                          & 855  & 857 &  911\\
 $m_{\tilde{e}_{L}}\,,m_{\tilde{\mu}_{L}}$             & 900  & 905 & 1014\\
 $m_{\tilde{\tau}_{2}}$                                & 860  & 863 &  919\\
 \hline \hline
 $m_{\neu}$                                            & 591 & 810  & 902 \\
 $m_{\smu}\,,m_{\sel}$                                 & 491 & 684  & 813 \\
 $m_{\staul}$                                         & 398 & 554  & 655  \\ 
 $m_{\snu}$                                           & 36.5 & 36.5 & 39.2\\
 \hline \hline
 $m_{h^{0}}$                                   & 124   &  125  &  125     \\
 $m_{A^{0}}$                                  & 1696   & 1800  & 1800     \\
 $\tan\beta$                                  & 11.18  & 20.00 & 30.00    \\
 $\mu$                                        & 1590   & 1200  &  930     \\ 
 \hline
 $\Omega_{\snu}{h^2}$                      & 0.1127  & 0.1128 &  0.1203   \\
 $A_{t}$                                   & -2374  & -2600 &  -2600   \\
 $A_{\nu}$                               & -2619    & -2619 & -2619  \\
 $|U^{\staul}_{L1}|$                     & 6.29$\times\,10^{-2}$ & 1.11$\times\,10^{-1}$ & 1.38$\times\,10^{-1}$ \\\hline
\end{tabular}
\caption{Benchmark points for studying the $m_{\neu}\,>\,m_{\smu\,,\sel}$ 
scenario.}
\label{BPcaseA}
\end{center}
\end{table}

\begin{table}[t]
 \begin{center}
\begin{tabular}{|c|c|c|c|}
\hline
Masses (in GeV) & BP4 & BP5 & BP6 \\\hline\hline
 $m_{\tilde{g}}$                          & 2190 & 2253  & 2253   \\
 $m_{\tilde{u}_{L}}\,,m_{\tilde{c}_{L}}$  & 1967 & 2322  & 2322  \\ 
 $m_{\tilde{u}_{R}}\,,m_{\tilde{c}_{R}}$  & 1885 & 2120  & 2120  \\
 $m_{\tilde{d}_{L}}\,,m_{\tilde{s}_{L}}$  & 1968 & 2323  & 2323  \\
 $m_{\tilde{d}_{R}}\,,m_{\tilde{s}_{R}}$  & 1877 & 2121  & 2121  \\ 
 $m_{\tilde{t}_{1}}$                      & 1182 & 1499  & 1500  \\
 $m_{\tilde{t}_{2}}$                      & 1730 & 2037  & 2039  \\
 $m_{\tilde{b}_{1}}$                      & 1666 & 1822  & 1827  \\
 $m_{\tilde{b}_{2}}$                      & 1705 & 2013  & 2017  \\ 
 \hline \hline
 $m_{\chi^{0}_{2}}$                       &  803 & 1017 & 1104  \\
 $m_{\chi^{\pm}_{1}}$                     &  803 & 1017 & 1103  \\
 $m_{\tilde{\nu}_{e_{L}}}\,,m_{\tilde{\nu}_{\mu_{L}}}$  & 894 & 1203 & 1204  \\
 $m_{\tilde{\nu}_{\tau_{L}}}$                     & 853 & 1103  & 1104  \\
 $m_{\tilde{e}_{L}}\,,m_{\tilde{\mu}_{L}}$        & 897 & 1206  & 1207  \\
 $m_{\tilde{\tau}_{2}}$                           & 859 & 1108  & 1112  \\
  \hline \hline
 $m_{\neu}$                  & 497 & 693  &  946  \\
 $m_{\smu}\,,m_{\sel}$       & 587 & 757  & 1006  \\
 $m_{\staul}$                & 421 & 599  &  831  \\
 $m_{\snu}$                  & 36.5 & 44.5  &  44.5    \\ 
 \hline \hline
 $m_{h^{0}}$        &  124  &  125  &  125  \\
 $m_{A^{0}}$        & 1696  & 1800  & 1800  \\
 $\tan\beta$        & 11.18 & 20.00 & 30.00 \\
 $\mu$              &  1590 &  1200 & 1200  \\
 \hline
 $\Omega_{\snu}{h^2}$     & 0.1127 & 0.1127 & 0.1112   \\
 $A_{t}$                  & -2375  & -2600  &  -2600   \\
 $A_{\nu}$                               & -2619    & -2619 & -2619  \\
 $|U^{\staul}_{L1}|$      & 6.49$\times\,10^{-2}$ & 5.58$\times\,10^{-1}$ & 1.33$\times\,10^{-1}$ \\\hline
\end{tabular}
\caption{Benchmark points for studying the $m_{\neu}\,<\,m_{\smu\,,\sel}$ 
scenario.}
\label{BPcaseB}
\end{center}
\end{table}

\subsection{The primary channel: one $\tau$-tagged jet}

Our primary signature is comprised of events with two $\staul/\mu$ tracks, two OSSF leptons (electrons 
and muons), one $\tau$-tagged jet along with $\met$ and it obeys the topology in Fig.~\ref{fig:signal}.
As the efficiency of tagging a hadronically  decaying $\tau$-lepton is below $100\%$, a statistically 
significant number of events end up with a single $\tau$-tagged jet. 
Thus, the final state having a single $\tau$-tagged jet calls for the use of the asymmetric $M_{T2}$ 
variable, which exhibits all the beneficial properties of the symmetric $M_{T2}$ variable but with the 
additional advantage discussed in Sec.~\ref{sec:strategy}. For the present work, we 
consider a $\tau$-tagging efficiency of $70\% \; (60\%)$ for the one- (three-) prong decay,  as discussed in Ref.~\cite{ATL-PHYS-PUB-2015-045}. The efficiency 
of mis-tagging a QCD jet as a tau-tagged jet has been chosen to be $\sim1\%\,-\,2\%$.

\begin{table}[h]
 \begin{center}
\begin{tabular}{|c|c|c|}
\hline
Cut Set    & \multicolumn{2}{c|}{$N_{s}$}  \\\hline
           & Case I   & Case II   \\ \hline
Cut Set A  & BP1: 73  & BP4: 45 \\
           & BP2: 26  & BP5: 11 \\
           & BP3: 10  & BP6: 2  \\ \hline \hline
Cut Set B & BP1: 79  & BP4: 48 \\
           & BP2: 31  & BP5: 12 \\
           & BP3: 12  & BP6: 2  \\ \hline           
\end{tabular}
\caption{Number of signal events, surviving all the cuts, at an integrated luminosity of 
$\mathcal{L}\,=\,3000$ fb$^{-1}$ for Case I ($m_{\neu}\,>\,m_{\smu\,,\sel}$ ) and Case II 
($m_{\neu}\,<\,m_{\smu\,,\sel}$) for the single $\tau$-tagged jet signature.}
\label{tab:caseABsig1}
\end{center}
\end{table}

\begin{figure}[h]
\begin{center}
\includegraphics[width=5.2cm,angle=270]{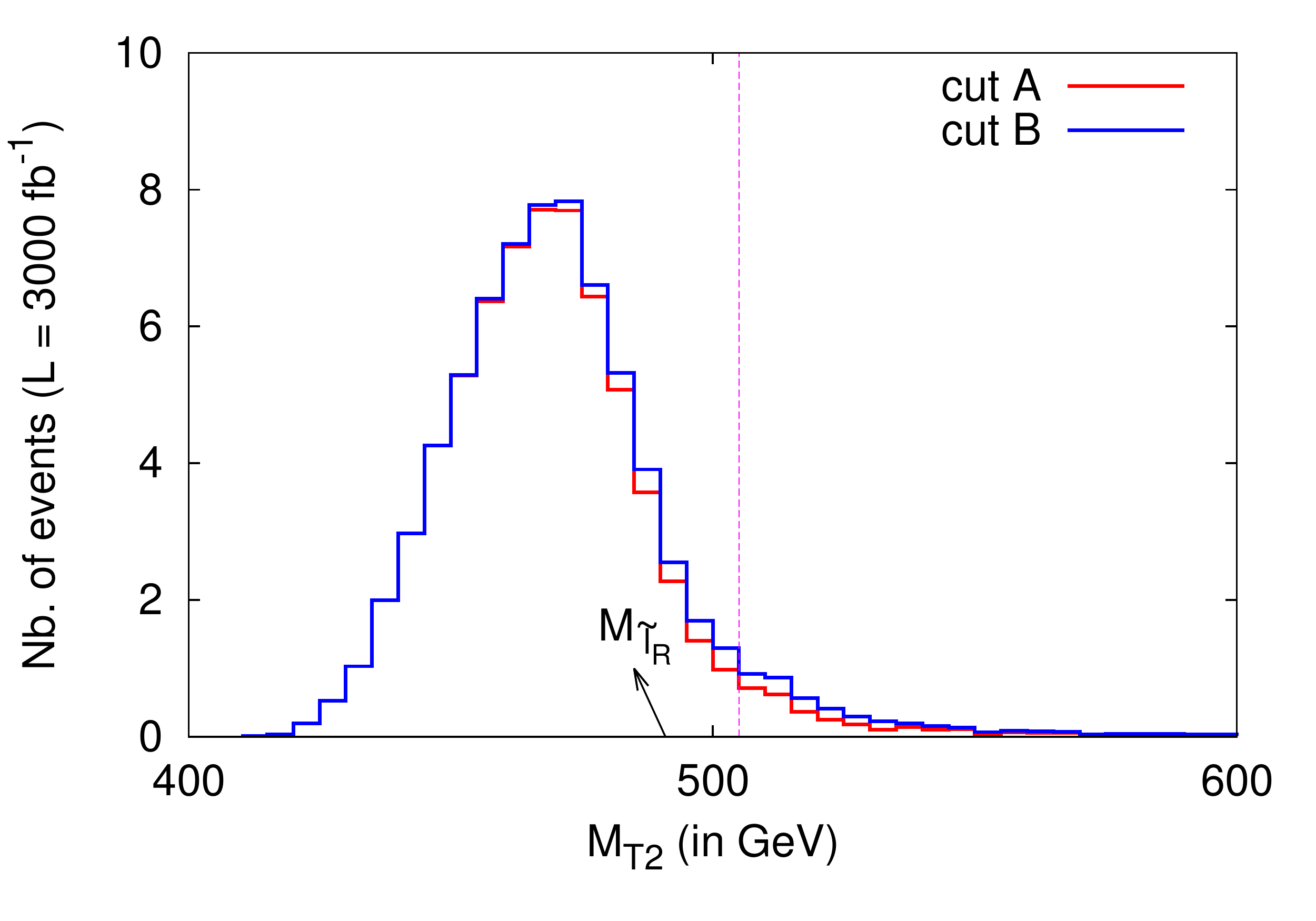}
\includegraphics[width=5.2cm,angle=270]{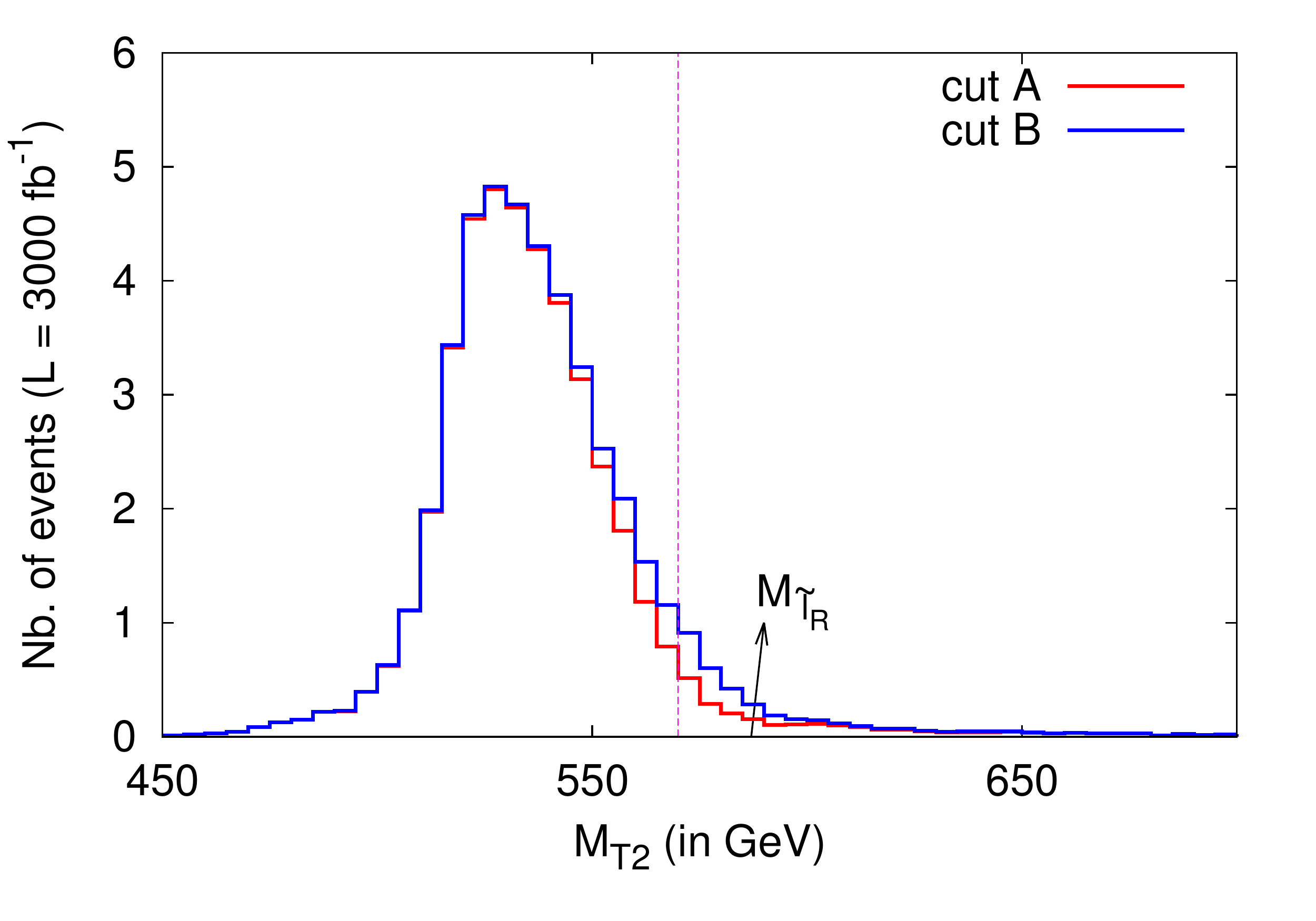}
\caption{$M_{T2}$-distributions for BP1 (left) and BP4 (right) corresponding to 
$m_{\neu}\,>\,m_{\smu\,,\sel}$ and $m_{\neu}\,<\,m_{\smu\,,\sel}$ respectively. The vertical 
dashed lines show the reconstructed slepton masses following our prescription while the arrow 
symbolises the actual slepton mass. The distributions are constructed after all cuts.}
\label{fig:MT2_dist}
\end{center}
\end{figure}

The number of signal events surviving all the cuts, at an integrated luminosity $\mathcal{L}\,=\,
3000$ fb$^{-1}$, are tabulated in Tab.~\ref{tab:caseABsig1} for both the mass hierarchies. The numbers 
include contributions from the process of interest, \textit{i.e.}, the Drell-Yan process 
as well as from the unwanted cascade topology. Both sets of cuts reduce the 
effect of the cascade contamination significantly. The $M_{T2}$-distributions for BP1 (case I) and BP4 (case II) are shown in 
Fig.~\ref{fig:MT2_dist} for the two sets of cuts which  differ only in their upper limit for the missing transverse momentum. 
One can clearly observe that cut set A lowers the number of events compared to cut set B, thereby 
improving the slepton mass reconstruction, by removing high $\met$ events which are mainly a 
manifestation of detector effects and longer distribution tails owing to the off-shell slepton
regime.  Finally, if one defines the 
end point of the $M_{T2}$ distribution to be the last bin that contains at least one signal event, 
then the slepton masses can be reconstructed with an accuracy of 5-10$\%$, at an integrated 
luminosity of 3000 fb$^{-1}$. Using this definition, the reconstructed (actual) slepton mass for BP1 
and BP4 are 505 (491) GeV and 570 (587) GeV respectively. The reconstructed (actual) masses are shown 
with  the  vertical dashed lines (arrows) in Fig.~\ref{fig:MT2_dist}.

Until now, we have focused on the number of signal events surviving all cuts. However, with the 
cut applied on the speed, $\beta$, of the two hardest muons as implemented in Ref.~\cite{ATLAS:2014fka}, we end up with hardly any background events. 
Indeed the total  SM background is reduced from  $\sim 21$ events for $\mathcal{L}\,=\,3000$ fb$^{-1}$ in the absence of the $\beta$-cut, to  
$\lesssim 1$ upon demanding $\beta \leqslant 0.98$ for the two hardest muons in each event.  Note that
to be realistic in our background modelling, we also take into account the possibility of QCD jets faking leptons. A flat mis-tagging rate of 0.5\% 
(0.1\%) is considered for $j \to e \; (\mu)$.

\subsection{Additional  channel : two $\tau$-tagged jet}

\begin{figure}[t]
\begin{center}
\includegraphics[width=5.2cm,angle=270]{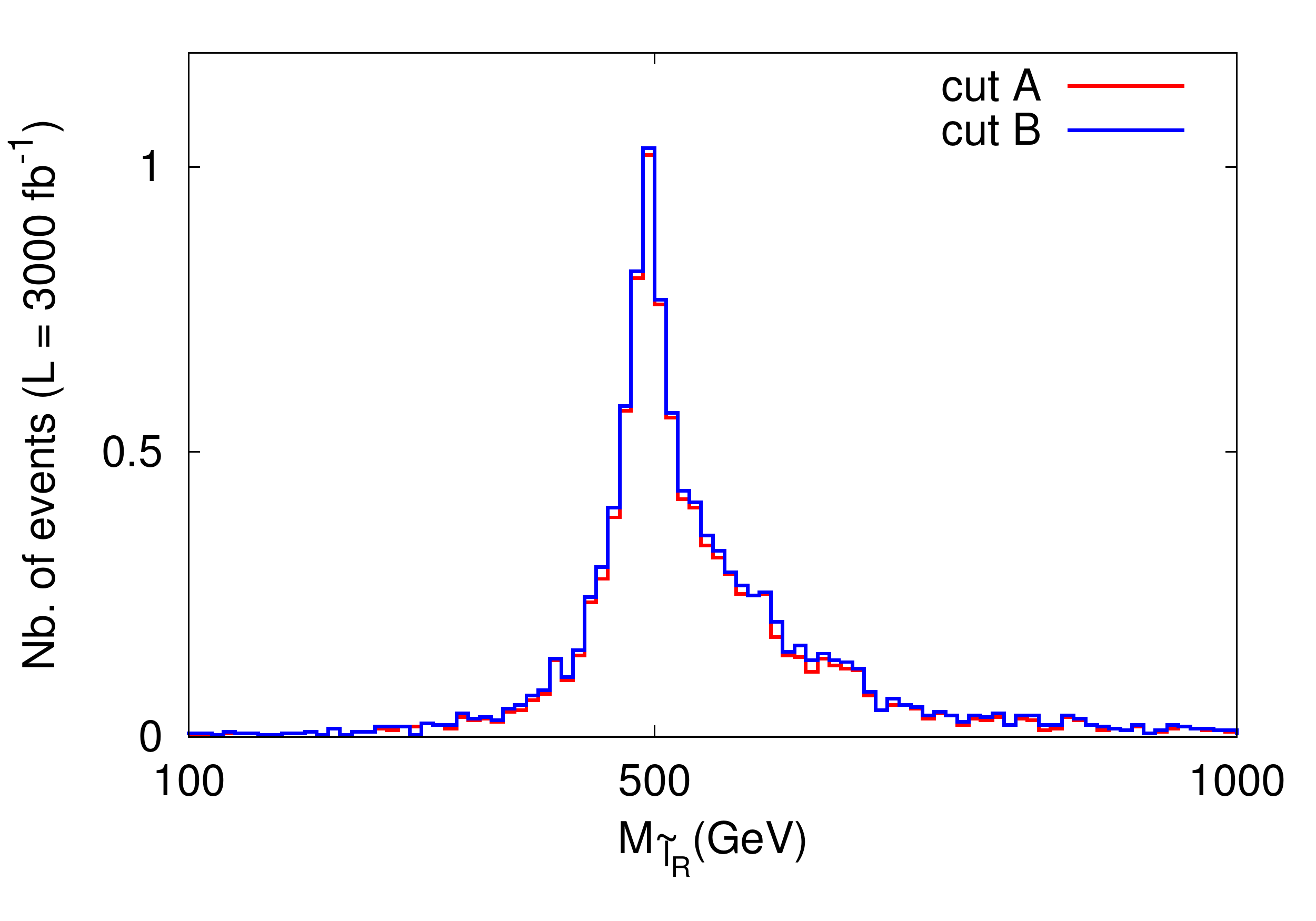}
\includegraphics[width=5.2cm,angle=270]{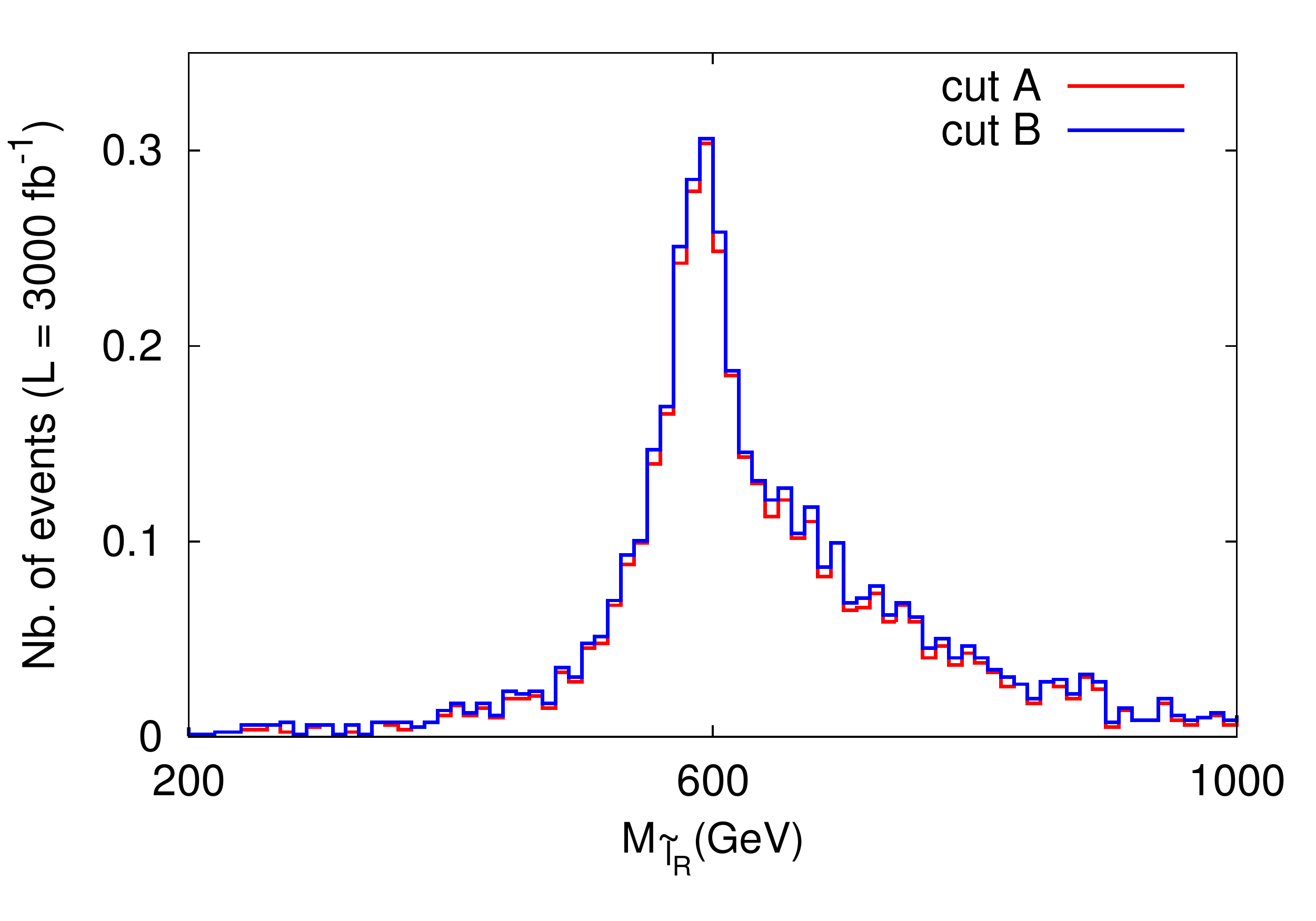}
\caption{Figure shows $m_{\smu/\sel}$ distributions for BP1 ($m_{\neu}\,>\,m_{\smu\,,\sel}$) on the 
left panel and BP4 ($m_{\neu}\,<\,m_{\smu\,,\sel}$) on the right panel. The distributions are 
constructed after all cuts.}
\label{fig:collinvmm}
\end{center}
\end{figure}

In this last segment, we focus on the signature comprising of 2 $\staul/\mu$ tracks, 2 OSSF leptons, 
2 $\tau$-tagged jets along with $\met$. This final state, however, suffers from a severe dearth of 
signals events owing to the double $\tau$-tag. We summarise the number of surviving signal events 
for the two mass hierarchies, in Tab.~\ref{tab:caseABsig2}.  Here, we use the collinear approximation 
in order to reconstruct the $\tau$s. This method is shown to work quite accurately for 
$p_{T}^{\tau_{j}}\,>\,40$ GeV. Fig.~\ref{fig:collinvmm} shows the final reconstruction of the 
slepton masses for BP1 and BP4. The reconstruction peaks agree with the actual masses within the 
percent level.

\begin{table}[h]
 \begin{center}
\begin{tabular}{|c|c|c|}
\hline
Cut Set    & \multicolumn{2}{c|}{$N_{s}$}  \\\hline
           & Case I     & Case II \\ \hline
Cut Set A & BP1: 12  & BP4: 11 \\
           & BP2: 7   & BP5: 3 \\
           & BP3: 2   & BP6: 1 \\ \hline \hline
Cut Set B & BP1: 13  & BP4: 12 \\
           & BP2: 9   & BP5: 3 \\
           & BP3: 3   & BP6: 1 \\ \hline           
\end{tabular}
\caption{Number of signal events, surviving all the cuts, at an integrated luminosity of 
$\mathcal{L}\,=\,3000$ fb$^{-1}$ for Case I ($m_{\neu}\,>\,m_{\smu\,,\sel}$ ) and Case II 
($m_{\neu}\,<\,m_{\smu\,,\sel}$) for the 2 $\staul/\mu$ + 2 $\tau$-tagged jet + 2 OSSF leptons + 
$\met$ final state.}
\label{tab:caseABsig2}
\end{center}
\end{table}

The number of background events for the double 
$\tau$-tagged scenario even before the implementation of the $\beta$ cut is more than an
order of magnitude smaller than its single $\tau$-tagged counterpart. With $\mathcal{L}\,=\,3000$ fb$^{-1}$, the number of background events is 
$\lesssim 1$. This is because, 
upon demanding two $\tau$ tags from the fake backgrounds ($t\bar{t}Z, ZZ$ and $ZW^+W^-$) 
with the small fake rates mentioned above, there are hardly any events which survive the
event selection. Moreover,  the real backgrounds, \textit{viz.}, $ZZZ$ 
$Zh$ have extremely small cross-sections  Furthermore, the $p_T$ requirement on the $\tau$-tagged jets reduces
the backgrounds further.  For consistency, we nevertheless  apply  the same cut on $\beta$ as in the previous case, moreover this cut  hardly affects the signal.
Even though the double-$\tau$-tagged events are ``background free'', the 
number of signal events is also very low. For most of the benchmark points 
 the number of signal events in the bin corresponding to 
actual slepton mass is less than one. Hence, although the 
two-$\tau$-tagged channel can in principle lead to a more accurate  reconstruction of the slepton masses  than the 
single $\tau$-tagged mode, this channel can only be useful for a future collider with much higher luminosities or higher energies than the HL-LHC.

To conclude this section, it is of utmost importance to reiterate that the lightest neutralino, 
$\chi^{0}_{1}$, can be reconstructed using the procedure outlined in Ref.~\cite{Biswas:2009zp} and the 
stau mass can be reconstructed using the method described in Ref.~\ref{eqn:staumass}. Hence, with the 
information on the reconstructed $\staul$-mass and the $\chi^{0}_{1}$-mass and the knowledge of 
reconstructing the right-handed slepton following the aforementioned procedures, one can 
straightforwardly disentangle the two mass-hierarchies, \textit{viz.}, $m_{\chi^{0}_{1}}<m_{\smu\,,\sel}$ and 
$m_{\smu\,,\sel}<m_{\chi^{0}_{1}}$.

\subsection{Detection prospects at MoEDAL}
Long-lived particles can also be looked for at the new and largely passive detector MoEDAL~\cite{Mavromatos:2016ykh, Acharya:2014nyr}.
It is composed of nuclear track detectors and is located at the Point 8 on the LHC ring. MoEDAL is designed to detect 
monopoles and massive stable charged particles.
Our model has a unique signature in terms of long-lived $\staul$s which
can be detected there, if their $\beta\,\lesssim\,0.5$. Although most of  the $\staul$s in the channels considered  do not satisfy this condition, see Fig.~\ref{fig:beta_dist}, at least one signal event is expected for all our benchmark points. We show in Tab.~\ref{tab:MoEDAL}, the number of events with single and double $\staul$s  expected at MoEDAL at an
integrated luminosity of $3000\,fb^{-1}$. For illustration, we have reported only those events with $p^{\staul}_{T}\,>\,10\,$
GeV and $\beta^{\staul}\,\lesssim\,0.5$. However, we have not taken into account, the angular orientations of these
long-lived particles and this may play a role in determining the final numbers. Although this signature will not provide additional information on the underlying SUSY spectrum, it will contribute to the validation of the long-lived stau scenario.

\begin{table}[h]
 \begin{center}
\begin{tabular}{|c|c|c|}
\hline
Benchmark points  & \multicolumn{2}{c|}{$N_{s}$}  \\\hline
           & 1 $\staul$  & 2 $\staul$\\ \hline
BP1        &      26          &   6            \\
BP2        &       7          &   2            \\
BP3        &       3          &   1            \\ \hline \hline
BP4        &      15          &   4            \\
BP5        &       4          &   1            \\
BP6        &       1          &   1            \\ \hline 
\end{tabular}
\caption{Number of events with 1 $\staul$ and 2 $\staul$, at an integrated luminosity of $\mathcal{L}\,=\,3000$ fb$^{-1}$ for Case I ($m_{\neu}\,>\,m_{\smu\,,\sel}$ ) and Case II ($m_{\neu}\,<\,m_{\smu\,,\sel}$) detectable at MoEDAL.}
\label{tab:MoEDAL}
\end{center}
\end{table}

\section{Summary and conclusions}\label{sec:summary}
A pMSSM scenario augmented with three families of right-chiral neutrino superfields has been assumed 
in this work. With only Dirac masses for neutrinos, and corresponding SUSY breaking scenario mass terms,
we have considered several benchmark points, with a right sneutrino as the LSP with the dominantly 
right-chiral $\staul$ serving as the NLSP. Owing to the smallness of the neutrino Yukawa coupling
(required by the neutrino oscillation data), the $\staul$s are fairly long-lived in the scale of 
colliders. Large $p_{T}$ and small $\beta$ of these long-lived particles make it easy to 
discriminate them from the SM backgrounds. We assumed two different hierarchical structures for the
masses of the weak-sector particles ($\neu$ and sleptons) in this work and have suggested a procedure 
for differentiating the two by reconstructing the slepton masses. We considered two possible signatures 
in each case, which differ only in the number of $\tau$-tagged jets identified in the final states. 
In case of the single $\tau$-tagged jet signal, the asymmetric $M_{T2}$ variable is found out to be 
a good kinematic variable while in the other case, the collinear approximation has been used to 
reconstruct the $\tau$s and thereby the sleptons. The latter method, even though cleaner, suffers
from a dearth in signal statistics and can only be used for future runs with higher luminosities and/or
centre of mass energies.
\section{Acknowledgements}\label{sec:Acknowledgements}
The authors thank AseshKrishna Datta, Shilpi Jain, Partha Konar, Mehedi Masud, Tanmoy Mondal and 
Michele Selvaggi for useful discussions at various stages of the work. The work of AG and BM is 
partially supported by funding available from the Department of Atomic Energy, Government of India, 
for the Regional Centre for Accelerator-based Particle Physics (RECAPP), Harish-Chandra Research 
Institute. SB is supported by a Durham Junior Research Fellowship COFUNDed between Durham University 
and the European Union under grant agreement number 609412. The work of GB is supported in part by the 
{\it Investissements d'avenir} Labex ENIGMASS.

\providecommand{\href}[2]{#2}
\addcontentsline{toc}{section}{References}
\bibliographystyle{JHEP}
\bibliography{refs}

\providecommand{\href}[2]{#2}\begingroup\raggedright\begin{thebibliography}{10}

\bibitem{Sofue:2000jx}
Y.~Sofue and V.~Rubin, \emph{{Rotation curves of spiral galaxies}},
  \href{https://doi.org/10.1146/annurev.astro.39.1.137}{\emph{Ann. Rev. Astron.
  Astrophys.} {\bfseries 39} (2001) 137--174},
  [\href{https://arxiv.org/abs/astro-ph/0010594}{{\ttfamily
  astro-ph/0010594}}].

\bibitem{Ade:2015xua}
{\scshape Planck} collaboration, P.~A.~R. Ade et~al., \emph{{Planck 2015
  results. XIII. Cosmological parameters}},
  \href{https://doi.org/10.1051/0004-6361/201525830}{\emph{Astron. Astrophys.}
  {\bfseries 594} (2016) A13},
  [\href{https://arxiv.org/abs/1502.01589}{{\ttfamily 1502.01589}}].

\bibitem{Markevitch:2001ri}
M.~Markevitch, A.~H. Gonzalez, L.~David, A.~Vikhlinin, S.~Murray, W.~Forman
  et~al., \emph{{A Textbook example of a bow shock in the merging galaxy
  cluster 1E0657-56}}, \href{https://doi.org/10.1086/339619}{\emph{Astrophys.
  J.} {\bfseries 567} (2002) L27},
  [\href{https://arxiv.org/abs/astro-ph/0110468}{{\ttfamily
  astro-ph/0110468}}].

\bibitem{Markevitch:2003at}
M.~Markevitch, A.~H. Gonzalez, D.~Clowe, A.~Vikhlinin, L.~David, W.~Forman
  et~al., \emph{{Direct constraints on the dark matter self-interaction
  cross-section from the merging galaxy cluster 1E0657-56}},
  \href{https://doi.org/10.1086/383178}{\emph{Astrophys. J.} {\bfseries 606}
  (2004) 819--824}, [\href{https://arxiv.org/abs/astro-ph/0309303}{{\ttfamily
  astro-ph/0309303}}].

\bibitem{Boehm:2003bt}
C.~Boehm, D.~Hooper, J.~Silk, M.~Casse and J.~Paul, \emph{{MeV dark matter: Has
  it been detected?}},
  \href{https://doi.org/10.1103/PhysRevLett.92.101301}{\emph{Phys. Rev. Lett.}
  {\bfseries 92} (2004) 101301},
  [\href{https://arxiv.org/abs/astro-ph/0309686}{{\ttfamily
  astro-ph/0309686}}].

\bibitem{Adriani:2008zr}
{\scshape PAMELA} collaboration, O.~Adriani et~al., \emph{{An anomalous
  positron abundance in cosmic rays with energies 1.5-100 GeV}},
  \href{https://doi.org/10.1038/nature07942}{\emph{Nature} {\bfseries 458}
  (2009) 607--609}, [\href{https://arxiv.org/abs/0810.4995}{{\ttfamily
  0810.4995}}].

\bibitem{Hooper:2010mq}
D.~Hooper and L.~Goodenough, \emph{{Dark Matter Annihilation in The Galactic
  Center As Seen by the Fermi Gamma Ray Space Telescope}},
  \href{https://doi.org/10.1016/j.physletb.2011.02.029}{\emph{Phys. Lett.}
  {\bfseries B697} (2011) 412--428},
  [\href{https://arxiv.org/abs/1010.2752}{{\ttfamily 1010.2752}}].

\bibitem{Accardo:2014lma}
{\scshape AMS} collaboration, L.~Accardo et~al., \emph{{High Statistics
  Measurement of the Positron Fraction in Primary Cosmic Rays of 0.5--500 GeV
  with the Alpha Magnetic Spectrometer on the International Space Station}},
  \href{https://doi.org/10.1103/PhysRevLett.113.121101}{\emph{Phys. Rev. Lett.}
  {\bfseries 113} (2014) 121101}.

\bibitem{Bulbul:2014sua}
E.~Bulbul, M.~Markevitch, A.~Foster, R.~K. Smith, M.~Loewenstein and S.~W.
  Randall, \emph{{Detection of An Unidentified Emission Line in the Stacked
  X-ray spectrum of Galaxy Clusters}},
  \href{https://doi.org/10.1088/0004-637X/789/1/13}{\emph{Astrophys. J.}
  {\bfseries 789} (2014) 13},
  [\href{https://arxiv.org/abs/1402.2301}{{\ttfamily 1402.2301}}].

\bibitem{Hooper:2008kg}
D.~Hooper, P.~Blasi and P.~D. Serpico, \emph{{Pulsars as the Sources of High
  Energy Cosmic Ray Positrons}},
  \href{https://doi.org/10.1088/1475-7516/2009/01/025}{\emph{JCAP} {\bfseries
  0901} (2009) 025}, [\href{https://arxiv.org/abs/0810.1527}{{\ttfamily
  0810.1527}}].

\bibitem{Boudaud:2014dta}
M.~Boudaud et~al., \emph{{A new look at the cosmic ray positron fraction}},
  \href{https://doi.org/10.1051/0004-6361/201425197}{\emph{Astron. Astrophys.}
  {\bfseries 575} (2015) A67},
  [\href{https://arxiv.org/abs/1410.3799}{{\ttfamily 1410.3799}}].

\bibitem{Undagoitia:2015gya}
T.~Marrodán~Undagoitia and L.~Rauch, \emph{{Dark matter direct-detection
  experiments}}, \href{https://doi.org/10.1088/0954-3899/43/1/013001}{\emph{J.
  Phys.} {\bfseries G43} (2016) 013001},
  [\href{https://arxiv.org/abs/1509.08767}{{\ttfamily 1509.08767}}].

\bibitem{Abercrombie:2015wmb}
D.~Abercrombie et~al., \emph{{Dark Matter Benchmark Models for Early LHC Run-2
  Searches: Report of the ATLAS/CMS Dark Matter Forum}},
  \href{https://arxiv.org/abs/1507.00966}{{\ttfamily 1507.00966}}.

\bibitem{Penning:2017tmb}
B.~Penning, \emph{{The Pursuit of Dark Matter at Colliders - An Overview}},
  \href{https://arxiv.org/abs/1712.01391}{{\ttfamily 1712.01391}}.

\bibitem{Buchmueller:2017qhf}
O.~Buchmueller, C.~Doglioni and L.~T. Wang, \emph{{Search for dark matter at
  colliders}}, \href{https://doi.org/10.1038/nphys4054}{\emph{Nature Phys.}
  {\bfseries 13} (2017) 217--223}.

\bibitem{Akerib:2018lyp}
{\scshape LUX-ZEPLIN} collaboration, D.~S. Akerib et~al., \emph{{Projected WIMP
  sensitivity of the LUX-ZEPLIN (LZ) dark matter experiment}},
  \href{https://arxiv.org/abs/1802.06039}{{\ttfamily 1802.06039}}.

\bibitem{Aprile:2017iyp}
{\scshape XENON} collaboration, E.~Aprile et~al., \emph{{First Dark Matter
  Search Results from the XENON1T Experiment}},
  \href{https://doi.org/10.1103/PhysRevLett.119.181301}{\emph{Phys. Rev. Lett.}
  {\bfseries 119} (2017) 181301},
  [\href{https://arxiv.org/abs/1705.06655}{{\ttfamily 1705.06655}}].

\bibitem{Essig:2011nj}
R.~Essig, J.~Mardon and T.~Volansky, \emph{{Direct Detection of Sub-GeV Dark
  Matter}}, \href{https://doi.org/10.1103/PhysRevD.85.076007}{\emph{Phys. Rev.}
  {\bfseries D85} (2012) 076007},
  [\href{https://arxiv.org/abs/1108.5383}{{\ttfamily 1108.5383}}].

\bibitem{Alexander:2016aln}
J.~Alexander et~al., \emph{{Dark Sectors 2016 Workshop: Community Report}},
  2016, \href{https://arxiv.org/abs/1608.08632}{{\ttfamily 1608.08632}},
  \href{http://inspirehep.net/record/1484628/files/arXiv:1608.08632.pdf}{http://inspirehep.net/record/1484628/files/arXiv:1608.08632.pdf}.

\bibitem{Crisler:2018gci}
{\scshape SENSEI} collaboration, M.~Crisler, R.~Essig, J.~Estrada,
  G.~Fernandez, J.~Tiffenberg, M.~Sofo~haro et~al., \emph{{SENSEI: First
  Direct-Detection Constraints on sub-GeV Dark Matter from a Surface Run}},
  \href{https://arxiv.org/abs/1804.00088}{{\ttfamily 1804.00088}}.

\bibitem{Feng:1997zr}
J.~L. Feng and T.~Moroi, \emph{{Tevatron signatures of longlived charged
  sleptons in gauge mediated supersymmetry breaking models}},
  \href{https://doi.org/10.1103/PhysRevD.58.035001}{\emph{Phys. Rev.}
  {\bfseries D58} (1998) 035001},
  [\href{https://arxiv.org/abs/hep-ph/9712499}{{\ttfamily hep-ph/9712499}}].

\bibitem{Bobrovskyi:2011vx}
S.~Bobrovskyi, W.~Buchmuller, J.~Hajer and J.~Schmidt, \emph{{Quasi-stable
  neutralinos at the LHC}},
  \href{https://doi.org/10.1007/JHEP09(2011)119}{\emph{JHEP} {\bfseries 09}
  (2011) 119}, [\href{https://arxiv.org/abs/1107.0926}{{\ttfamily 1107.0926}}].

\bibitem{Brandenburg:2005he}
A.~Brandenburg, L.~Covi, K.~Hamaguchi, L.~Roszkowski and F.~D. Steffen,
  \emph{{Signatures of axinos and gravitinos at colliders}},
  \href{https://doi.org/10.1016/j.physletb.2005.04.072}{\emph{Phys. Lett.}
  {\bfseries B617} (2005) 99--111},
  [\href{https://arxiv.org/abs/hep-ph/0501287}{{\ttfamily hep-ph/0501287}}].

\bibitem{Belanger:2011ny}
G.~Belanger, S.~Kraml and A.~Lessa, \emph{{Light Sneutrino Dark Matter at the
  LHC}}, \href{https://doi.org/10.1007/JHEP07(2011)083}{\emph{JHEP} {\bfseries
  07} (2011) 083}, [\href{https://arxiv.org/abs/1105.4878}{{\ttfamily
  1105.4878}}].

\bibitem{Dumont:2012ee}
B.~Dumont, G.~Belanger, S.~Fichet, S.~Kraml and T.~Schwetz, \emph{{Mixed
  sneutrino dark matter in light of the 2011 XENON and LHC results}},
  \href{https://doi.org/10.1088/1475-7516/2012/09/013}{\emph{JCAP} {\bfseries
  1209} (2012) 013}, [\href{https://arxiv.org/abs/1206.1521}{{\ttfamily
  1206.1521}}].

\bibitem{Banerjee:2013fga}
S.~Banerjee, P.~S.~B. Dev, S.~Mondal, B.~Mukhopadhyaya and S.~Roy,
  \emph{{Invisible Higgs Decay in a Supersymmetric Inverse Seesaw Model with
  Light Sneutrino Dark Matter}},
  \href{https://doi.org/10.1007/JHEP10(2013)221}{\emph{JHEP} {\bfseries 10}
  (2013) 221}, [\href{https://arxiv.org/abs/1306.2143}{{\ttfamily 1306.2143}}].

\bibitem{Arina:2015uea}
C.~Arina, M.~E.~C. Catalan, S.~Kraml, S.~Kulkarni and U.~Laa,
  \emph{{Constraints on sneutrino dark matter from LHC Run 1}},
  \href{https://doi.org/10.1007/JHEP05(2015)142}{\emph{JHEP} {\bfseries 05}
  (2015) 142}, [\href{https://arxiv.org/abs/1503.02960}{{\ttfamily
  1503.02960}}].

\bibitem{Abdallah:2018gjj}
W.~Abdallah, A.~Hammad, A.~Kasem and S.~Khalil, \emph{{Long-Lived BLSSM
  Particles at the LHC}},  \href{https://arxiv.org/abs/1804.09778}{{\ttfamily
  1804.09778}}.

\bibitem{Asaka:2005cn}
T.~Asaka, K.~Ishiwata and T.~Moroi, \emph{{Right-handed sneutrino as cold dark
  matter}}, \href{https://doi.org/10.1103/PhysRevD.73.051301}{\emph{Phys. Rev.}
  {\bfseries D73} (2006) 051301},
  [\href{https://arxiv.org/abs/hep-ph/0512118}{{\ttfamily hep-ph/0512118}}].

\bibitem{Asaka:2006fs}
T.~Asaka, K.~Ishiwata and T.~Moroi, \emph{{Right-handed sneutrino as cold dark
  matter of the universe}},
  \href{https://doi.org/10.1103/PhysRevD.75.065001}{\emph{Phys. Rev.}
  {\bfseries D75} (2007) 065001},
  [\href{https://arxiv.org/abs/hep-ph/0612211}{{\ttfamily hep-ph/0612211}}].

\bibitem{Falk:1994es}
T.~Falk, K.~A. Olive and M.~Srednicki, \emph{{Heavy sneutrinos as dark
  matter}}, \href{https://doi.org/10.1016/0370-2693(94)90639-4}{\emph{Phys.
  Lett.} {\bfseries B339} (1994) 248--251},
  [\href{https://arxiv.org/abs/hep-ph/9409270}{{\ttfamily hep-ph/9409270}}].

\bibitem{Arina:2007tm}
C.~Arina and N.~Fornengo, \emph{{Sneutrino cold dark matter, a new analysis:
  Relic abundance and detection rates}},
  \href{https://doi.org/10.1088/1126-6708/2007/11/029}{\emph{JHEP} {\bfseries
  11} (2007) 029}, [\href{https://arxiv.org/abs/0709.4477}{{\ttfamily
  0709.4477}}].

\bibitem{Tan:2016zwf}
{\scshape PandaX-II} collaboration, A.~Tan et~al., \emph{{Dark Matter Results
  from First 98.7 Days of Data from the PandaX-II Experiment}},
  \href{https://doi.org/10.1103/PhysRevLett.117.121303}{\emph{Phys. Rev. Lett.}
  {\bfseries 117} (2016) 121303},
  [\href{https://arxiv.org/abs/1607.07400}{{\ttfamily 1607.07400}}].

\bibitem{Akerib:2016vxi}
{\scshape LUX} collaboration, D.~S. Akerib et~al., \emph{{Results from a search
  for dark matter in the complete LUX exposure}},
  \href{https://doi.org/10.1103/PhysRevLett.118.021303}{\emph{Phys. Rev. Lett.}
  {\bfseries 118} (2017) 021303},
  [\href{https://arxiv.org/abs/1608.07648}{{\ttfamily 1608.07648}}].

\bibitem{Chala:2017jgg}
M.~Chala, A.~Delgado, G.~Nardini and M.~Quiros, \emph{{A light sneutrino
  rescues the light stop}},
  \href{https://doi.org/10.1007/JHEP04(2017)097}{\emph{JHEP} {\bfseries 04}
  (2017) 097}, [\href{https://arxiv.org/abs/1702.07359}{{\ttfamily
  1702.07359}}].

\bibitem{Capozzi:2018ubv}
F.~Capozzi, E.~Lisi, A.~Marrone and A.~Palazzo, \emph{{Current unknowns in the
  three neutrino framework}},
  \href{https://arxiv.org/abs/1804.09678}{{\ttfamily 1804.09678}}.

\bibitem{Martin:1997ns}
S.~P. Martin, \emph{{A Supersymmetry primer}},
  \href{https://arxiv.org/abs/hep-ph/9709356}{{\ttfamily hep-ph/9709356}}.

\bibitem{Banerjee:2016uyt}
S.~Banerjee, G.~B\'{e}langer, B.~Mukhopadhyaya and P.~D. Serpico,
  \emph{{Signatures of sneutrino dark matter in an extension of the CMSSM}},
  \href{https://doi.org/10.1007/JHEP07(2016)095}{\emph{JHEP} {\bfseries 07}
  (2016) 095}, [\href{https://arxiv.org/abs/1603.08834}{{\ttfamily
  1603.08834}}].

\bibitem{Evans:2016zau}
J.~A. Evans and J.~Shelton, \emph{{Long-Lived Staus and Displaced Leptons at
  the LHC}}, \href{https://doi.org/10.1007/JHEP04(2016)056}{\emph{JHEP}
  {\bfseries 04} (2016) 056},
  [\href{https://arxiv.org/abs/1601.01326}{{\ttfamily 1601.01326}}].

\bibitem{Heisig:2011dr}
J.~Heisig and J.~Kersten, \emph{{Production of long-lived staus in the
  Drell-Yan process}},
  \href{https://doi.org/10.1103/PhysRevD.84.115009}{\emph{Phys. Rev.}
  {\bfseries D84} (2011) 115009},
  [\href{https://arxiv.org/abs/1106.0764}{{\ttfamily 1106.0764}}].

\bibitem{Hinchliffe:1998ys}
I.~Hinchliffe and F.~E. Paige, \emph{{Measurements in gauge mediated SUSY
  breaking models at CERN LHC}},
  \href{https://doi.org/10.1103/PhysRevD.60.095002}{\emph{Phys. Rev.}
  {\bfseries D60} (1999) 095002},
  [\href{https://arxiv.org/abs/hep-ph/9812233}{{\ttfamily hep-ph/9812233}}].

\bibitem{Biswas:2009zp}
S.~Biswas and B.~Mukhopadhyaya, \emph{{Neutralino reconstruction in
  supersymmetry with long-lived staus}},
  \href{https://doi.org/10.1103/PhysRevD.79.115009}{\emph{Phys. Rev.}
  {\bfseries D79} (2009) 115009},
  [\href{https://arxiv.org/abs/0902.4349}{{\ttfamily 0902.4349}}].

\bibitem{Biswas:2009rba}
S.~Biswas and B.~Mukhopadhyaya, \emph{{Chargino reconstruction in supersymmetry
  with long-lived staus}},
  \href{https://doi.org/10.1103/PhysRevD.81.015003}{\emph{Phys. Rev.}
  {\bfseries D81} (2010) 015003},
  [\href{https://arxiv.org/abs/0910.3446}{{\ttfamily 0910.3446}}].

\bibitem{Chatterjee:2016rjo}
A.~Chatterjee, N.~Chakrabarty and B.~Mukhopadhyaya, \emph{{Same-sign trileptons
  as a signal of sneutrino lightest supersymmetric partlcle}},
  \href{https://doi.org/10.1016/j.physletb.2015.12.074}{\emph{Phys. Lett.}
  {\bfseries B754} (2016) 14--17},
  [\href{https://arxiv.org/abs/1411.7226}{{\ttfamily 1411.7226}}].

\bibitem{Biswas:2010cd}
S.~Biswas, \emph{{Reconstruction of the left-chiral tau-sneutrino in
  supersymmetry with a right-sneutrino as the lightest supersymmetric
  particle}}, \href{https://doi.org/10.1103/PhysRevD.82.075020}{\emph{Phys.
  Rev.} {\bfseries D82} (2010) 075020},
  [\href{https://arxiv.org/abs/1002.4395}{{\ttfamily 1002.4395}}].

\bibitem{Gupta:2007ui}
S.~K. Gupta, B.~Mukhopadhyaya and S.~K. Rai, \emph{{Right-chiral sneutrinos and
  long-lived staus: Event characteristics at the large hadron collider}},
  \href{https://doi.org/10.1103/PhysRevD.75.075007}{\emph{Phys. Rev.}
  {\bfseries D75} (2007) 075007},
  [\href{https://arxiv.org/abs/hep-ph/0701063}{{\ttfamily hep-ph/0701063}}].

\bibitem{Lester:1999tx}
C.~G. Lester and D.~J. Summers, \emph{{Measuring masses of semiinvisibly
  decaying particles pair produced at hadron colliders}},
  \href{https://doi.org/10.1016/S0370-2693(99)00945-4}{\emph{Phys. Lett.}
  {\bfseries B463} (1999) 99--103},
  [\href{https://arxiv.org/abs/hep-ph/9906349}{{\ttfamily hep-ph/9906349}}].

\bibitem{Barr:2003rg}
A.~Barr, C.~Lester and P.~Stephens, \emph{{m(T2): The Truth behind the
  glamour}}, \href{https://doi.org/10.1088/0954-3899/29/10/304}{\emph{J. Phys.}
  {\bfseries G29} (2003) 2343--2363},
  [\href{https://arxiv.org/abs/hep-ph/0304226}{{\ttfamily hep-ph/0304226}}].

\bibitem{Lattanzi:2017ubx}
M.~Lattanzi and M.~Gerbino, \emph{{Status of neutrino properties and future
  prospects - Cosmological and astrophysical constraints}},
  \href{https://doi.org/10.3389/fphy.2017.00070}{\emph{Front.in Phys.}
  {\bfseries 5} (2018) 70}, [\href{https://arxiv.org/abs/1712.07109}{{\ttfamily
  1712.07109}}].

\bibitem{Djouadi:1998di}
{\scshape MSSM Working Group} collaboration, A.~Djouadi et~al., \emph{{The
  Minimal supersymmetric standard model: Group summary report}},
  \href{https://arxiv.org/abs/hep-ph/9901246}{{\ttfamily hep-ph/9901246}}.

\bibitem{Belanger:2014vza}
G.~Bélanger, F.~Boudjema, A.~Pukhov and A.~Semenov, \emph{{micrOMEGAs4.1: two
  dark matter candidates}},
  \href{https://doi.org/10.1016/j.cpc.2015.03.003}{\emph{Comput. Phys. Commun.}
  {\bfseries 192} (2015) 322--329},
  [\href{https://arxiv.org/abs/1407.6129}{{\ttfamily 1407.6129}}].

\bibitem{Hall:2009bx}
L.~J. Hall, K.~Jedamzik, J.~March-Russell and S.~M. West, \emph{{Freeze-In
  Production of FIMP Dark Matter}},
  \href{https://doi.org/10.1007/JHEP03(2010)080}{\emph{JHEP} {\bfseries 03}
  (2010) 080}, [\href{https://arxiv.org/abs/0911.1120}{{\ttfamily 0911.1120}}].

\bibitem{McDonald:2001vt}
J.~McDonald, \emph{{Thermally generated gauge singlet scalars as
  selfinteracting dark matter}},
  \href{https://doi.org/10.1103/PhysRevLett.88.091304}{\emph{Phys. Rev. Lett.}
  {\bfseries 88} (2002) 091304},
  [\href{https://arxiv.org/abs/hep-ph/0106249}{{\ttfamily hep-ph/0106249}}].

\bibitem{Aaboud:2018wps}
{\scshape ATLAS} collaboration, M.~Aaboud et~al., \emph{{Measurement of the
  Higgs boson mass in the $H\rightarrow ZZ^* \rightarrow 4\ell$ and $H
  \rightarrow \gamma\gamma$ channels with $\sqrt{s}=13$ TeV $pp$ collisions
  using the ATLAS detector}},
  \href{https://arxiv.org/abs/1806.00242}{{\ttfamily 1806.00242}}.

\bibitem{Aad:2015zhl}
{\scshape ATLAS, CMS} collaboration, G.~Aad et~al., \emph{{Combined Measurement
  of the Higgs Boson Mass in $pp$ Collisions at $\sqrt{s}=7$ and 8 TeV with the
  ATLAS and CMS Experiments}},
  \href{https://doi.org/10.1103/PhysRevLett.114.191803}{\emph{Phys. Rev. Lett.}
  {\bfseries 114} (2015) 191803},
  [\href{https://arxiv.org/abs/1503.07589}{{\ttfamily 1503.07589}}].

\bibitem{Khachatryan:2014jba}
{\scshape CMS} collaboration, V.~Khachatryan et~al., \emph{{Precise
  determination of the mass of the Higgs boson and tests of compatibility of
  its couplings with the standard model predictions using proton collisions at
  7 and 8 $\,\text {TeV}$}},
  \href{https://doi.org/10.1140/epjc/s10052-015-3351-7}{\emph{Eur. Phys. J.}
  {\bfseries C75} (2015) 212},
  [\href{https://arxiv.org/abs/1412.8662}{{\ttfamily 1412.8662}}].

\bibitem{Aad:2015gba}
{\scshape ATLAS} collaboration, G.~Aad et~al., \emph{{Measurements of the Higgs
  boson production and decay rates and coupling strengths using pp collision
  data at $\sqrt{s}=7$ and 8 TeV in the ATLAS experiment}},
  \href{https://doi.org/10.1140/epjc/s10052-015-3769-y}{\emph{Eur. Phys. J.}
  {\bfseries C76} (2016) 6},
  [\href{https://arxiv.org/abs/1507.04548}{{\ttfamily 1507.04548}}].

\bibitem{Bernon:2015hsa}
J.~Bernon and B.~Dumont, \emph{{Lilith: a tool for constraining new physics
  from Higgs measurements}},
  \href{https://doi.org/10.1140/epjc/s10052-015-3645-9}{\emph{Eur. Phys. J.}
  {\bfseries C75} (2015) 440},
  [\href{https://arxiv.org/abs/1502.04138}{{\ttfamily 1502.04138}}].

\bibitem{Bechtle:2013wla}
P.~Bechtle, O.~Brein, S.~Heinemeyer, O.~Stål, T.~Stefaniak, G.~Weiglein et~al.,
  \emph{{$\mathsf{HiggsBounds}-4$: Improved Tests of Extended Higgs Sectors
  against Exclusion Bounds from LEP, the Tevatron and the LHC}},
  \href{https://doi.org/10.1140/epjc/s10052-013-2693-2}{\emph{Eur. Phys. J.}
  {\bfseries C74} (2014) 2693},
  [\href{https://arxiv.org/abs/1311.0055}{{\ttfamily 1311.0055}}].

\bibitem{Kawasaki:2004qu}
M.~Kawasaki, K.~Kohri and T.~Moroi, \emph{{Big-Bang nucleosynthesis and
  hadronic decay of long-lived massive particles}},
  \href{https://doi.org/10.1103/PhysRevD.71.083502}{\emph{Phys. Rev.}
  {\bfseries D71} (2005) 083502},
  [\href{https://arxiv.org/abs/astro-ph/0408426}{{\ttfamily
  astro-ph/0408426}}].

\bibitem{Kawasaki:2017bqm}
M.~Kawasaki, K.~Kohri, T.~Moroi and Y.~Takaesu, \emph{{Revisiting Big-Bang
  Nucleosynthesis Constraints on Long-Lived Decaying Particles}},
  \href{https://doi.org/10.1103/PhysRevD.97.023502}{\emph{Phys. Rev.}
  {\bfseries D97} (2018) 023502},
  [\href{https://arxiv.org/abs/1709.01211}{{\ttfamily 1709.01211}}].

\bibitem{CMS-PAS-EXO-16-036}
{\scshape CMS Collaboration} collaboration, \emph{{Search for heavy stable
  charged particles with $12.9~\mathrm{fb}^{-1}$ of 2016 data}},  Tech. Rep.
  CMS-PAS-EXO-16-036, CERN, Geneva, 2016.

\bibitem{Aaboud:2017vwy}
{\scshape ATLAS} collaboration, M.~Aaboud et~al., \emph{{Search for squarks and
  gluinos in final states with jets and missing transverse momentum using 36
  fb$^{-1}$ of $\sqrt{s}$=13 TeV $pp$ collision data with the ATLAS detector}},
   \href{https://arxiv.org/abs/1712.02332}{{\ttfamily 1712.02332}}.

\bibitem{Sirunyan:2017kqq}
{\scshape CMS} collaboration, A.~M. Sirunyan et~al., \emph{{Search for new
  phenomena with the $M_{\mathrm {T2}}$ variable in the all-hadronic final
  state produced in proton--proton collisions at $\sqrt{s} = 13$ $\,\text
  {TeV}$}}, \href{https://doi.org/10.1140/epjc/s10052-017-5267-x}{\emph{Eur.
  Phys. J.} {\bfseries C77} (2017) 710},
  [\href{https://arxiv.org/abs/1705.04650}{{\ttfamily 1705.04650}}].

\bibitem{ATLAS:2014fka}
{\scshape ATLAS} collaboration, G.~Aad et~al., \emph{{Searches for heavy
  long-lived charged particles with the ATLAS detector in proton-proton
  collisions at $ \sqrt{s}=8 $ TeV}},
  \href{https://doi.org/10.1007/JHEP01(2015)068}{\emph{JHEP} {\bfseries 01}
  (2015) 068}, [\href{https://arxiv.org/abs/1411.6795}{{\ttfamily 1411.6795}}].

\bibitem{Rainwater:1998kj}
D.~L. Rainwater, D.~Zeppenfeld and K.~Hagiwara, \emph{{Searching for
  $H\to\tau^+\tau^-$ in weak boson fusion at the CERN LHC}},
  \href{https://doi.org/10.1103/PhysRevD.59.014037}{\emph{Phys. Rev.}
  {\bfseries D59} (1998) 014037},
  [\href{https://arxiv.org/abs/hep-ph/9808468}{{\ttfamily hep-ph/9808468}}].

\bibitem{Konar:2009qr}
P.~Konar, K.~Kong, K.~T. Matchev and M.~Park, \emph{{Dark Matter Particle
  Spectroscopy at the LHC: Generalizing M(T2) to Asymmetric Event Topologies}},
  \href{https://doi.org/10.1007/JHEP04(2010)086}{\emph{JHEP} {\bfseries 04}
  (2010) 086}, [\href{https://arxiv.org/abs/0911.4126}{{\ttfamily 0911.4126}}].

\bibitem{Mangano:2006rw}
M.~L. Mangano, M.~Moretti, F.~Piccinini and M.~Treccani, \emph{{Matching matrix
  elements and shower evolution for top-quark production in hadronic
  collisions}},
  \href{https://doi.org/10.1088/1126-6708/2007/01/013}{\emph{JHEP} {\bfseries
  01} (2007) 013}, [\href{https://arxiv.org/abs/hep-ph/0611129}{{\ttfamily
  hep-ph/0611129}}].

\bibitem{Chatrchyan:2013oca}
{\scshape CMS} collaboration, S.~Chatrchyan et~al., \emph{{Searches for
  long-lived charged particles in pp collisions at $\sqrt{s}$=7 and 8 TeV}},
  \href{https://doi.org/10.1007/JHEP07(2013)122}{\emph{JHEP} {\bfseries 07}
  (2013) 122}, [\href{https://arxiv.org/abs/1305.0491}{{\ttfamily 1305.0491}}].

\bibitem{Alwall:2014hca}
J.~Alwall, R.~Frederix, S.~Frixione, V.~Hirschi, F.~Maltoni, O.~Mattelaer
  et~al., \emph{{The automated computation of tree-level and next-to-leading
  order differential cross sections, and their matching to parton shower
  simulations}}, \href{https://doi.org/10.1007/JHEP07(2014)079}{\emph{JHEP}
  {\bfseries 07} (2014) 079},
  [\href{https://arxiv.org/abs/1405.0301}{{\ttfamily 1405.0301}}].

\bibitem{Sjostrand:2014zea}
T.~Sjöstrand, S.~Ask, J.~R. Christiansen, R.~Corke, N.~Desai, P.~Ilten et~al.,
  \emph{{An Introduction to PYTHIA 8.2}},
  \href{https://doi.org/10.1016/j.cpc.2015.01.024}{\emph{Comput. Phys. Commun.}
  {\bfseries 191} (2015) 159--177},
  [\href{https://arxiv.org/abs/1410.3012}{{\ttfamily 1410.3012}}].

\bibitem{Cacciari:2008gp}
M.~Cacciari, G.~P. Salam and G.~Soyez, \emph{{The Anti-k(t) jet clustering
  algorithm}}, \href{https://doi.org/10.1088/1126-6708/2008/04/063}{\emph{JHEP}
  {\bfseries 04} (2008) 063},
  [\href{https://arxiv.org/abs/0802.1189}{{\ttfamily 0802.1189}}].

\bibitem{Cacciari:2011ma}
M.~Cacciari, G.~P. Salam and G.~Soyez, \emph{{FastJet User Manual}},
  \href{https://doi.org/10.1140/epjc/s10052-012-1896-2}{\emph{Eur. Phys. J.}
  {\bfseries C72} (2012) 1896},
  [\href{https://arxiv.org/abs/1111.6097}{{\ttfamily 1111.6097}}].

\bibitem{deFavereau:2013fsa}
{\scshape DELPHES 3} collaboration, J.~de~Favereau, C.~Delaere, P.~Demin,
  A.~Giammanco, V.~Lemaître, A.~Mertens et~al., \emph{{DELPHES 3, A modular
  framework for fast simulation of a generic collider experiment}},
  \href{https://doi.org/10.1007/JHEP02(2014)057}{\emph{JHEP} {\bfseries 02}
  (2014) 057}, [\href{https://arxiv.org/abs/1307.6346}{{\ttfamily 1307.6346}}].

\bibitem{Ball:2012cx}
R.~D. Ball et~al., \emph{{Parton distributions with LHC data}},
  \href{https://doi.org/10.1016/j.nuclphysb.2012.10.003}{\emph{Nucl. Phys.}
  {\bfseries B867} (2013) 244--289},
  [\href{https://arxiv.org/abs/1207.1303}{{\ttfamily 1207.1303}}].

\bibitem{Beenakker:1996ed}
W.~Beenakker, R.~Hopker and M.~Spira, \emph{{PROSPINO: A Program for the
  production of supersymmetric particles in next-to-leading order QCD}},
  \href{https://arxiv.org/abs/hep-ph/9611232}{{\ttfamily hep-ph/9611232}}.

\bibitem{ATL-PHYS-PUB-2015-045}
\emph{{Reconstruction, Energy Calibration, and Identification of Hadronically
  Decaying Tau Leptons in the ATLAS Experiment for Run-2 of the LHC}},  Tech.
  Rep. ATL-PHYS-PUB-2015-045, CERN, Geneva, Nov, 2015.

\bibitem{Mavromatos:2016ykh}
{\scshape MoEDAL} collaboration, N.~E. Mavromatos and V.~A. Mitsou,
  \emph{{Physics reach of MoEDAL at LHC: magnetic monopoles, supersymmetry and
  beyond}}, \href{https://doi.org/10.1051/epjconf/201716404001}{\emph{EPJ Web
  Conf.} {\bfseries 164} (2017) 04001},
  [\href{https://arxiv.org/abs/1612.07012}{{\ttfamily 1612.07012}}].

\bibitem{Acharya:2014nyr}
{\scshape MoEDAL} collaboration, B.~Acharya et~al., \emph{{The Physics
  Programme Of The MoEDAL Experiment At The LHC}},
  \href{https://doi.org/10.1142/S0217751X14300506}{\emph{Int. J. Mod. Phys.}
  {\bfseries A29} (2014) 1430050},
  [\href{https://arxiv.org/abs/1405.7662}{{\ttfamily 1405.7662}}].

\end{thebibliography}\endgroup

\end{document}